\documentclass[sigconf]{acmart}

\usepackage{url}
\usepackage{listings}
\usepackage{color,soul}
\usepackage{amsmath}
\usepackage{pgfplots}
\usepackage{subcaption}
\usepackage{multicol}
\usepackage{xfrac}
\usepackage{enumitem}
\usepackage[binary-units]{siunitx}

\setlist[itemize]{noitemsep, topsep=0pt}
\usepackage{orcidlink}
\usepgfplotslibrary{colorbrewer}
\usepgfplotslibrary{fillbetween}

\usetikzlibrary{intersections}
\usetikzlibrary{patterns}
\usetikzlibrary{pgfplots.groupplots}

\pgfplotsset{compat=1.18}
\usepgfplotslibrary{units}

\definecolor{s1}{RGB}{228, 26, 28}
\definecolor{s2}{RGB}{55, 126, 184}
\definecolor{s3}{RGB}{77, 175, 74}
\definecolor{s4}{RGB}{152, 78, 163}
\definecolor{s5}{RGB}{255, 127, 0}
\definecolor{s6}{RGB}{25, 127, 0}
\definecolor{s7}{RGB}{25, 127, 55}
\definecolor{s8}{RGB}{255, 127, 45}
\definecolor{s9}{RGB}{85, 127, 65}
\definecolor{s10}{RGB}{0, 0, 255}
\definecolor{s11}{RGB}{0, 0, 0}
\definecolor{s12}{RGB}{255, 0, 0}

\pgfplotscreateplotcyclelist{set2}{
    s1, style={fill=s1}\\
    s2, style={fill=s2}\\
    s8, style={fill=s8}\\
    s9, style={fill=s9}\\
}

\copyrightyear{2024}
\acmYear{2024}
\setcopyright{acmlicensed}\acmConference[HPQCI '24]{Workshop on High
Performance and Quantum Computing Integration}{June 3--4, 2024}{Pisa, Italy}
\acmBooktitle{Workshop on High Performance and Quantum Computing
Integration (HPQCI '24), June 3--4, 2024, Pisa, Italy}
\acmDOI{10.1145/3659996.3660036}
\acmISBN{979-8-4007-0643-1/24/06}

\lstdefinestyle{python}{
    backgroundcolor=\color{gray!10},
    basicstyle=\ttfamily\footnotesize,
    breaklines=true,
    captionpos=b,
    commentstyle=\color{gray},
    keywordstyle=\color{blue},
    stringstyle=\color{orange},
    frame=tb,
    language=Python,
    numbers=left,
    numbersep=5pt,
    numberstyle=\tiny\color{gray},
    showspaces=false,
    showstringspaces=false,
    tabsize=2,
    columns=flexible
}

\AtBeginDocument{%
	\providecommand\BibTeX{{%
			\normalfont B\kern-0.5em{\scshape i\kern-0.25em b}\kern-0.8em\TeX}}}

\definecolor{punct}{HTML}{CC3311}
\definecolor{background}{HTML}{EEEEEE}
\definecolor{delim}{HTML}{0077BB}
\definecolor{numb}{HTML}{EE7733}
\definecolor{grey}{HTML}{BBBBBB}
\definecolor{lightgrey}{HTML}{DDDDDD}
\definecolor{lightestgrey}{HTML}{EEEEEE}

\lstdefinelanguage{json}{
    basicstyle=\footnotesize\ttfamily,
    numbers=left,
    numberstyle=\footnotesize,
    stepnumber=1,
    numbersep=4pt,
    showstringspaces=false,
    breaklines=true,
    frame=lines,
    backgroundcolor=\color{background},
    literate=
     *{0}{{{\color{numb}0}}}{1}
      {1}{{{\color{numb}1}}}{1}
      {2}{{{\color{numb}2}}}{1}
      {3}{{{\color{numb}3}}}{1}
      {4}{{{\color{numb}4}}}{1}
      {5}{{{\color{numb}5}}}{1}
      {6}{{{\color{numb}6}}}{1}
      {7}{{{\color{numb}7}}}{1}
      {8}{{{\color{numb}8}}}{1}
      {9}{{{\color{numb}9}}}{1}
      {:}{{{\color{punct}{:}}}}{1}
      {,}{{{\color{punct}{,}}}}{1}
      {\{}{{{\color{delim}{\{}}}}{1}
      {\}}{{{\color{delim}{\}}}}}{1}
      {[}{{{\color{delim}{[}}}}{1}
      {]}{{{\color{delim}{]}}}}{1},
}

\begin{document}

\title{Quantum Mini-Apps: A Framework for Developing and Benchmarking Quantum-HPC Applications}

\author{Nishant Saurabh\orcidlink{0000-0002-1926-4693}}
\email{n.saurabh@uu.nl}
\affiliation{
  \institution{Utrecht University}
  \city{Utrecht}
  \country{NL}
}

\author{Pradeep Mantha\orcidlink{0000-0003-2664-7737}}
\email{pradeepm66@gmail.com}
\affiliation{
  \institution{Ludwig Maximilian University}
  \city{Munich}
  \country{Germany}
}

\author{Florian J. Kiwit\orcidlink{0009-0000-4065-1535}}
\email{f.kiwit@campus.lmu.de}
\affiliation{
  \institution{BMW Group, Ludwig Maximilian University}
  \city{Munich}
  \country{Germany}
}

\author{Shantenu Jha\orcidlink{0000-0002-5040-026X}}
\email{shantenu.jha@rutgers.edu}
\affiliation{
  \institution{Rutgers University}
  \city{New Brunswick, New Jersey}
  \country{USA}
}

\author{Andre Luckow\orcidlink{0000-0002-1225-4062}}
\email{andre.luckow@ifi.lmu.de}%

\affiliation{
  \institution{BMW Group, Ludwig Maximilian University}
  \city{Munich}
  \country{Germany}
}

\begin{abstract}
With the increasing maturity and scale of quantum hardware and its integration into HPC systems, there is a need to develop robust techniques for developing, characterizing, and benchmarking quantum-HPC applications and middleware systems. This requires a better understanding of interaction, coupling, and common execution patterns between quantum and classical workload tasks and components. This paper identifies six quantum-HPC execution motifs—recurring execution patterns characterized by distinct coupling and interaction modes. These motifs provide the basis for a suite of quantum mini-apps -- simplified application prototypes that encapsulate essential characteristics of production systems. To support these developments, we introduce a mini-app framework that offers the necessary abstractions for creating and executing mini-apps across heterogeneous quantum-HPC infrastructure, making it a valuable tool for performance characterizations and middleware development.

\end{abstract}



\maketitle

\section{Introduction}

Quantum algorithms might outperform classical algorithms in the future for certain tasks by requiring fewer computational steps~\cite{dalzell2023quantum}. Quantum algorithms are particularly relevant for high-performance computing (HPC) applications across various scientific and industrial domains~\cite{alexeev2023quantumcentric, bayerstadler2021industry}. For example, in materials science, quantum computing offers novel ways to explore molecular interactions and properties at the atomic level~\cite{alexeev2023quantumcentric}.

However, current quantum computers are limited in their capabilities. Noisy Intermediate Scale Quantum Computers (NISQ)~\cite{Preskill2018quantumcomputingin} are particularly limited regarding qubit quality, number, and gate depth. Thus, increasingly, efforts have shifted from NISQ to Fault-tolerant Quantum Computers (FTQC)~\cite{NAP26850}. NISQ and FTQC require developing hybrid systems that merge quantum and classical computing capabilities. For example, NISQ variational algorithms and FTQC's quantum error correction routines combine quantum with classical computation. Specifically, quantum processing units (QPUs) must be effectively integrated at the hardware and software level into HPC systems, forming effective \emph{quantum-HPC systems}~\cite{alexeev2023quantumcentric, elsharkawy2023integration}. 

As quantum-HPC systems and applications grow in complexity and scale, there is a need to design robust techniques for their benchmarking, characterization, and development.  However, specific characteristics of quantum algorithms, e.g., the interactions of quantum and classical components, the integration of algorithms in workflows, and execution, are poorly understood. Further, diverse quantum hardware systems, e.\,g., superconducting, ion-trap, and neutral atom-based systems, make it challenging to optimize applications and workflows. For example, these different qubit modalities have inherently different connectivity, noise, error, and decoherence properties. On the software level, a significant challenge lies in integrating quantum primitives into algorithms that combine quantum and classical components. Such algorithms eventually form part of larger workflows, including data conversion, encoding, pre-, and post-processing. 

Given the complex nature of quantum systems, a quantum-HPC middleware system will be crucial for the efficient development of quantum-HPC applications. The software stack for quantum-HPC is and will remain fragmented, specializing in specific hardware, algorithms, and applications.  Many crucial questions must be addressed when developing such systems: What are common application scenarios? How are quantum and classical components coupled? How can the diverse software and hardware stack be better integrated, enabling higher abstractions while achieving optimal performance? What are common execution patterns? How can these components be mapped to classical and quantum processing elements, including quantum simulations? How can the scalability and performance of quantum applications be characterized and optimized? How can quantum applications be conceptualized and prototyped efficiently? 

To address the aforementioned challenges effectively, we propose the \emph{quantum mini-app} framework to conceptualize quantum solutions. Mini-apps~\cite{9006530} are simplified versions of complex computational applications replicating these systems' key components and performance characteristics. Existing mini-apps~\cite{pilot-streaming,workflow_miniapps_jha2024} typically focus on classical systems and do not fully encompass the unique properties of quantum-HPC applications and middleware systems. The design of our quantum mini-apps is based on a comprehensive analysis of current algorithms and applications, from which we systematically extract and categorize recurring execution patterns into so-called execution motifs. These execution motifs capture essential properties of quantum-HPC applications, e.\,g., the interaction and coupling between classical and quantum components. Further, we design and develop a mini-app framework that simplifies the creation of mini-apps and their integration into existing quantum HPC, cloud systems, and simulators.

This paper's main contributions and structure are as follows. 
We survey different application scenarios and execution patterns to develop six motifs in section~\ref{sec:motif}. The \emph{quantum mini-app} framework allows the rapid iteration of application prototypes and motifs and is presented in Section~\ref{sec:mini_apps}. We demonstrate the framework using an example mini-app for circuit execution in Section~\ref{sec:eval}. Section~\ref{sec:related} discusses the related works on mini-apps and quantum benchmarks. Section~\ref{sec:conclusion} summarizes the paper and discusses possible future work.

\section{Quantum-HPC Integration Types and Execution Motifs}
\label{sec:quantum_miniapps}

\begin{figure}[t]
  \includegraphics[width=0.40\textwidth]{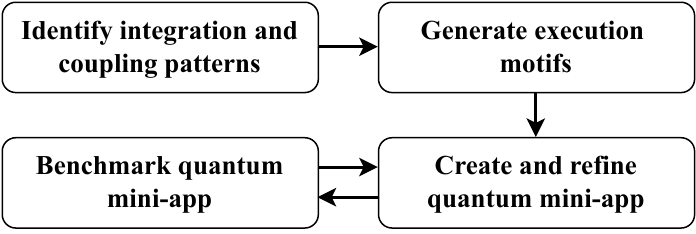}
  \caption{Quantum-HPC integration types, execution motifs, and mini-app conceptual stages\label{fig:qc_miniapps}}
\end{figure}

This section lays the foundation for the design of the quantum mini-apps. After describing the mini-app development stages and the main quantum-HPC integration types, we introduce the concept of an execution motif, i.e., re-occurring patterns in integrating classical and quantum components. 

\subsection{Stages of Developing Quantum Mini-Apps}

Figure~\ref{sec:quantum_miniapps} shows the conceptual stages of a quantum mini-app. It involves three stages as follows:

\emph{(i)} The first stage aims to understand the interaction between classical HPC and quantum components and identify their \emph{integration} and \emph{coupling patterns}~\cite{saurabh2023conceptual}. Identifying quantum-HPC integration strategies can help estimate communication frequency and time-sensitivity between diverse task components and benchmark their orchestrated resource usage, i.e., the proportion of time utilized by computations on HPC and quantum hardware and simulators.

\emph{(ii)} The second stage involves \emph{generating execution motifs}, which are fine-grained execution patterns, i.\,e., a unique workflow configuration that combines different modes of coupling between quantum and classical components and captures a particular relevant pattern (adapted from~\cite{jha2024aihpc}). Execution motifs can help benchmark and optimize task mapping and distribution across \emph{processing elements} (PEs), where a PE is a CPU core, a GPU, or other computational units within specialized processors or accelerators capable of executing computing tasks.
    
\emph{(ii)} The third stage focuses on \emph{creating quantum mini-app} that encapsulates one or more high-level execution motifs emulating unique quantum-HPC workload patterns based on their interaction types within and outside QPU coherence time. Developed mini-apps can further utilize one or more quantum benchmarks~\cite{lubinski2024quantum, kiwit2024benchmarking} to characterize the orchestrated execution patterns and further evaluate and \emph{refine} the mini-app. Integrating quantum mini-apps to such benchmarks is part of our future work.
\label{sec:motif}
Next, we discuss three integration types and six execution motifs for quantum-HPC systems.

\subsection{Quantum-HPC Integration Types}
\label{sec:app_integration_types}
Understanding how the various application components interact is crucial to developing representative mini-apps and their performance characterization. Quantum and classical coupling are the interactions between application components within and outside the quantum system. In~\cite{saurabh2023conceptual}, we identified three integration patterns.

\emph{(i) HPC-for-Quantum:} This pattern emphasizes using classical HPC techniques to enhance quantum computing. It includes optimizing low-level system operations such as input/output, dynamic circuits~\cite{PhysRevLett.127.100501}, error mitigation~\cite{ella2023quantumclassical}, and effectively utilizing quantum processing units (QPUs). This approach is crucial for real-time interactions between HPC and quantum tasks, focusing on low-level circuit developers and leveraging HPC techniques like parallelization and high-performance networking.

\emph{(ii) Quantum-in-HPC:} This pattern integrates a medium-coupled quantum component with a classical HPC component. Unlike the HPC-for-Quantum pattern, classical and quantum computation integration does not occur in real-time. This pattern typically encompasses scenarios where quantum tasks are orchestrated by a classical HPC system, with the quantum component providing acceleration for specific types of tasks. Examples of Quantum-in-HPC~\cite{saurabh2023conceptual} pattern involves accelerating ground state energy estimation in a molecular system or linear equation solving~\cite{vqls,vqa}.

\emph{(iii) Quantum-about-HPC:} A quantum-enhanced kernel is integrated into a broader quantum-classical workflow in this integration pattern. The quantum component is added without significantly modifying the HPC application structure. This approach involves looser coupling than other patterns~\cite{saurabh2023conceptual}, and the main application control typically resides in the classical system. The quantum component acts as a specialized tool within the workflow, often requiring additional input or output processing to be effective~\cite{https://doi.org/10.48550/arxiv.2101.06250,generative_molecule_design_2022}.

\begin{table*}[t]
\centering
\small
\caption{Mapping quantum-HPC execution motifs based on the interaction and coupling patterns between quantum and classical tasks and components. Further, it investigates the middleware capabilities required to support this motif.}
\vspace*{-0.2cm}
\label{tab:hpc_quantum_patterns}
\renewcommand{\arraystretch}{1} %
\resizebox{\textwidth}{!}{
\small
\begin{tabular}{|c|c|c|c|c|c|}
\hline
\cline{1-5}
{{\textbf{Motif}}} & \multicolumn{1}{c|}{{\textbf{Interaction}}}&\multicolumn{1}{c|}{{\textbf{Coupling}}}&\multicolumn{1}{c|}{{\textbf{Example}}}&\multicolumn{1}{c|}{{\textbf{Middleware Capabilities}}}
\\
\hline
\hline
\textit{\textbf{Circuit}}&Concurrent &Loosely coupled,&Qiskit Estimator/ &Workload\\
\textit{\textbf{Execution}}& computation for&homogeneous &Sampler, Qiskit Aer &management across\\
\textit{}&different parameters &tasks&with Dask~\cite{QiskitAerParallel} &heterogeneous resources\\
\hline
\hline
\textit{\textbf{Distributed}}&Concurrent computation& Tightly coupled & Large-scale quantum&Integration\\
\textit{\textbf{State Vector}}& of state vector updates&static and& simulations, e.\,g.,&with HPC technologies, e.\,g.,\\
\textit{\textbf{Simulation}}& \& its synchronization& homogeneous&cuQuantum &MPI, cuQuantum, ROCm \\
\textit{}&across all tasks& tasks&~\cite{cuquantum}&\\
\hline
\hline
\textit{\textbf{Circuit}}&Concurrent execution&Medium-to-tight coupling&Qiskit Circuit&Optimal partitioning
\\
\textit{\textbf{Cutting}}& \& reconstruction of& within/outside coherence time,&Knitting Toolbox~\cite{circuit-knitting-toolbox},&\& placements of tasks/cuts \\
\textit{}&tasks across &task heterogeneity depends&Pennylane Circuit&on available (simulated) QPUs \\
\textit{}&QPUs.&on circuit cutting algorithm&Cutting~\cite{PennyLaneQCut}& for balanced workload.\\
\hline
\hline
\textit{\textbf{Error}}&Loosely coupled &&Qiskit&Dynamic allocation \& adaptation \\
\textit{\textbf{Mitigation}}&execution of multiple &Loosely &Runtime Error& to right mix of classical \\
\textit{}&circuit variants on PEs&coupled&Mitigation& \& quantum resources\\
\textit{}& \& results aggregation&&~\cite{IBMQuantumConfigureErrorMitigation}& 
\\
\hline
\hline
\textit{\textbf{Variational}}&Concurrent execution&Coupling outside coherence&QAOA~\cite{farhi2014quantum}, VQE~\cite{vqe}, &Collocate \& balance\\
\textit{\textbf{Quantum}}&of classical and quantum resources \& & window for heterogeneous&VQE IS~\cite{vqa_is} &quantum and classical \\
\textit{\textbf{Algorithms (VQA)}}&Quantum& tasks, e.\,g., interleaved &QGAN~\cite{riofrio2023performance}, QuGen~\cite{QutacQuantumQuGen} &resources\\
\textit{}&components& ML \& QC parts sharing GPU&&\\
\hline
\hline
\textit{\textbf{Multistage}}&Encapsulated stages \& &Heterogeneous \& &QML workflow& Optimized resource estimation
\\
\textit{\textbf{Pipelines}}& its contained transitions&varying resource demands&MD workflows~\cite{cranganore2024paving}&for pipeline/individual stages,\\
\textit{}& with control \& data flow&between stages&&e.\,g., for dynamic resource pooling\\
\hline
\end{tabular}
}
\end{table*}

\subsection{Execution Motifs}
\label{sec:execution_motifs}

We adapt the concept of execution motifs~\cite{Brewer2024AICoupledHPC} to the quantum-HPC domain, describing recurring patterns of interaction and coupling between classical and quantum components. Specifically, we propose six representative motifs covering three integration types: HPC-for-Quantum, Quantum-in-HPC, and Quantum-about-HPC.  We describe quantum-HPC execution motifs focusing primarily on interaction and coupling characteristics:

\emph{(i) Interaction:} describe how quantum and HPC components interact. These include the movement and management of data, the control and coordination of processes, and the capacity for real-time interactions between quantum and HPC elements. Quantum and classical tasks can be run concurrently in full or partial concurrent or sequential modes.

\emph{(ii) Coupling} refers to the characteristics of the employed execution strategies, e.g., simultaneous operations (concurrency), the diversity of components (heterogeneity), and the variability in how the workflow is structured (dynamism). The type of coupling between quantum and classical tasks is essential for quantum-HPC applications, e.g., within or outside the coherence time~\cite{alexeev2023quantumcentric}. This work considers three types of coupling between quantum and classical tasks. Tasks with strict time-sensitive interactions within QPU coherence time fall into the tightly-coupled category, while time-sensitive interactions of tasks outside the QPU coherence time are medium-coupled. The third coupling type, i.\,e., loosely-coupled, refers to tasks with infrequent interactions and involves seamless integration of interdependent classical and quantum task components.

Table~\ref{tab:hpc_quantum_patterns} provides a detailed mapping of interaction and coupling patterns for each motif below and their benefits in designing and refining quantum-HPC systems. Further, we investigate the middleware capabilities required to support these.

\emph{(i) Circuit execution:} This motif can be found in the HPC-for-Quantum integration type. Utilizing loosely and medium-coupled task parallelism is important when optimizing the execution of quantum circuits using parallelism and multiple processing elements (PEs) (both classical and quantum), e.g., for sampling and computing estimation value.  Another example is parameterized circuits, which require executing the same circuit using different parameters. Different PEs can be used to improve the estimation, e.g., of expectation values.  Another form of parallelism is the partition of different terms of a Hamiltonian across multiple PEs. Abstraction and tools for supporting this parallelism exist on different levels, e.g., Qiskit Aer provides a multi-processing and Dask executor on the backend device level~\cite{QiskitAerParallel}, Qiskit Serverless on the middleware level. 
  
\emph{(ii) Distributed state vector simulation} is another motif found in the HPC-for-Quantum category. It utilizes multiple processing elements, i.\,e., cores, nodes, and GPU, to benchmark the computational and memory needs of quantum simulations by partitioning and distributing the state vector, i.\,e., the state of a quantum system. In this motif, coupling is tight and occurs between classical tasks. Updates to the state vector are done by multiplying a unitary matrix. This computation is conducted concurrently. Depending on the type of operation, only local or non-local qubits, i.\,e., qubits placed on different processing elements, can be affected. Operations on local qubits can be performed without data exchange, while non-local or global qubits may require significant data movement. Thus, MPI is commonly used to facilitate the communication between tasks. Examples of distributed state vectors include QULAC (CPU/GPU)~\cite{imamura2022mpiqulacs} and cuQuantum's cuStateVec (GPU)~\cite{cuquantum}. Further, different programming frameworks utilize cuQuantum to provide a distributed state vector simulation, e.\,g., Pennylane~\cite{bergholm2018pennylane} and Qiskit~\cite{qiskit_1}. 
    
\emph{(iii) Circuit cutting} emerged as a technique to mitigate the availability of a limited number of qubits on current QPUs by partitioning circuits onto multiple QPUs~\cite{Peng_2020, circuit-knitting-toolbox}. For example, Tomesh et al.~\cite{10313799} utilize circuit cutting for implementing a distributed optimization algorithm for the maximum independent set problem. Different cutting techniques have been proposed. Coupling between tasks can occur via a classical network (or in the future via a quantum network). Depending on the type of circuit cutting, the workload tasks can be distributed across QPU, and classical resources (e.g., simulated QPUs on CPU/GPU). For example, Clifford-based cutting \cite{10.1145/3579371.3589352} partitions circuits into efficiently simulatable Clifford circuits that can be run on classical resources, and the remaining non-Clifford circuits run on QPUs.
    
\emph{(iv) Error Mitigation:} This motif could simulate error mitigation techniques in quantum computing, providing insights into the effectiveness of various error mitigation strategies on different quantum hardware. Many strategies rely on executing circuit ensembles on both multiple QPUs and simulators. For example, Ravi et\,al.~\cite{ravi2022boosting} utilize classically simulated Clifford canaries to improve the fidelity of the quantum circuit execution.
    
\emph{(v) Variational Quantum Algorithms (VQA)}, such as the Variational Quantum Eigensolver (VQE)~\cite{vqe} and the Quantum Approximate Optimization Algorithm (QAOA)~\cite{farhi2014quantum}, interleave (and potentially concurrently execute) quantum and classical components to solve optimization, machine learning, linear algebra, factoring or chemistry problems~\cite{vqa}. 

Depending on the algorithm, the amount of classical/quantum computing and interleaving between both can vary significantly. For example, in vanilla VQE and QAOA, classical computing primarily concerns data preparation, post-processing, and optimization. More advanced algorithms, e.\,g., VQE with information sharing~\cite{vqa_is} and quantum machine learning (QML), utilize more complex or multiple cost functions requiring more classical computing. QGANs, e.\,g., utilize a complex GAN cost function and can benefit from gradient-based optimizers, thus leading to increased classical computational requirements~\cite{riofrio2023performance}. Accelerators, e.\,g., GPUs, for parallel gradient computation and backpropagation, can significantly enhance performance~\cite{kiwit2023application}. In these cases, sharing resources for ML and quantum computation, such as the GPU memory for quantum simulation and ML back-propagation tasks, is often necessary. 

Usually, these stages of variational algorithms, i.\,e., state preparation, quantum circuit execution, measurement, expectation value computation, and classical optimization, are executed sequentially. Execution motifs, such as the circuit execution and distributed state vector motif, can be applied to individual stages, e.\,g., for optimizing circuit execution for parallel sampling and expectation value computation. Further, classical and quantum algorithms can be interleaved and executed concurrently in advanced scenarios. For example, recursive QAOA algorithms like QIRO~\cite{finžgar2024quantuminformed} could be adapted to run variational algorithms concurrently with a classical optimization algorithm concurrently exchanging information between both frequently. The described VQE with information sharing~\cite{vqa_is} solves multiple related variational circuits in parallels and utilizes multiple Bayesian optimizers that incorporate the measurement results from these problems.

\emph{(vi) Multistage pipeline} motif comprises multiple stages, e.\,g., data preparation, training, and inference. The pipeline pattern is typically used with other motifs, e.g., state vector simulation or variational algorithms. An example is chemistry applications requiring multiple stages, e.\,g., reading molecular data as input, setting up the molecular system, running algorithms for molecular energy level calculations, and conducting post-processing steps. Another example is a machine learning pipeline, requiring multiple stages, e.g., data compression, training, and inference~\cite{jobst2023efficient}. Cranganorea et\,al.'s~\cite{cranganore2024paving} molecular dynamic workflow comprising multiple stages that can be enhanced with quantum algorithms, e.\,g., it proposes the utilization of the CSWAP algorithm for distance calculation computation and VQE for computation of the eigenvalues.

In the future, we expect more complex, generalized workflows beyond simple pipelines. In such workflows, directed acyclic graphs (DAG) will play a crucial role in modeling and representing the structure of a quantum workflow by describing sequences of tasks (nodes) and their dependencies (edges). New integration patterns between classical and quantum machine learning are emerging, e.\,g., the transformer-based quantum eigensolver (GPT-QE)~\cite{nakaji2024generative}. GPT-QE utilizes a transformer architecture based on GPT-2 to generate sequences of indices for unitary operations from a predefined operator set, effectively combining classical computational resources (CPUs and GPUs) with quantum processors (QPUs). The application encompasses multiple stages: dataset creation, pre-training, training, and generation.

\begin{figure*}[t]
  \includegraphics[width=\textwidth]{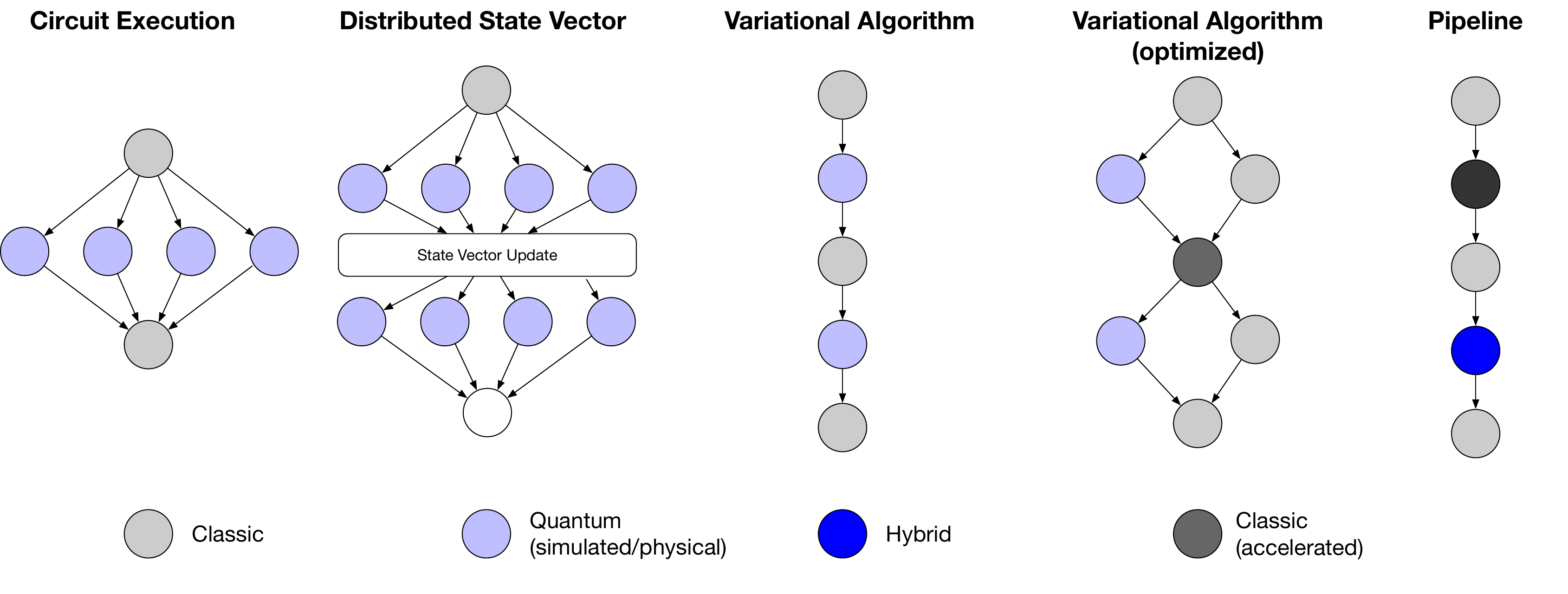}
  \caption{Quantum-HPC core motifs: circuit execution, distributed state vector, variational algorithms (standard and optimized), and the pipeline motif: quantum-HPC applications can exhibit complex task structures and increasingly integrate optimized HPC capabilities, e.\,g., GPU acceleration and other forms of parallelization.\label{fig:qc_motifs}}
\end{figure*}

\subsection{Discussion}

Quantum-HPC applications feature increasingly complex task topology that progressively utilize more advanced HPC capabilities, including GPU acceleration and diverse parallelization strategies (e.\,g., data and task parallelism). Figure~\ref{fig:qc_motifs} visualizes the interaction between quantum and classical components and couplings for four key execution motifs discussed in section~\ref{sec:execution_motifs}, i.\,e., the circuit execution, distributed state vector simulation, variational algorithms (in two variants), and the pipeline motif. 

For quantum simulations, both loosely-coupled ensemble and tightly-coupled MPI-based parallelism are critical. Loosely-coupled parallelism can be exploited when sampling circuits or computing expectation values. Tightly coupled distributed state vector simulations that require multi-node systems are necessary for large qubit counts. GPU-accelerated infrastructure can enhance the simulation's complexity, quality, and runtime for these scenarios.

As quantum algorithms and applications mature, more sophisticated algorithms emerge. For example, we highlight two variants of the variational algorithm motif. The first variant demonstrates a sequential approach, commonly found in proof-of-principle implementations, focusing on the iterative optimization process between quantum circuit evaluation and classical optimization. In contrast, the \emph{optimized} variant utilizes various types of parallelism. For example, quantum and classical tasks can run concurrently, e.\,g., recursive QAOA algorithms like QIRO~\cite{finžgar2024quantuminformed} could continuously run a classical optimization model that is steered using the results of the quantum circuits. Further, QML applications can leverage GPUs for faster gradient computation and updates.

The pipeline motif represents multi-stage workflows where quantum and classical tasks are interlinked to process and analyze data. Each stage might utilize specialized parallelism, e.\,g., data-parallelism is important for data preparation and encoding.

The middleware required to support various quantum-HPC execution motifs must facilitate seamless integration, workload management, and resource allocation across heterogeneous computing infrastructures. Essential capabilities include handling classical and quantum tasks, optimizing task partitioning and placement to balance the resource mix, dynamically managing resource demands, and integrating error correction and mitigation routines.

\section{Quantum Mini-Apps}
\label{sec:mini_apps}
This section discusses quantum mini-app design objectives and proposes three mini-app examples. Next, we briefly discuss our quantum mini-app framework architecture which is still in the early stages of development, and demonstrate it using a circuit execution motif. The quantum mini-app framework is developed in Python, and the source code is available on Github~\cite{QuantumMiniApp}.

\subsection{Design Objectives}
Quantum applications and systems are still in their infancy. With the increasing availability of quantum hardware, there is a need to develop benchmarks for evaluating the performance of quantum-HPC systems. It involves creating scalable and production-ready quantum software, robust performance characterization and benchmarking tools, and methodologies. We postulate that quantum mini-apps can be an efficient tool for researchers and developers to design, evaluate, refine, and understand the interaction between quantum hardware and software. Additionally, it will enable the definition of concrete application requirements and the drawing of broader conclusions on the design of abstraction and capabilities of quantum-HPC systems. We identify three distinct design objectives of quantum mini-apps as follows.

\emph{(i) Bridging Quantum Hardware and Software:} Mini-apps can provide a better understanding of mitigating the limitations of current quantum hardware, such as coherence time, gate fidelity, and qubit count. By characterizing these constraints using relevant application workloads, developers can effectively manage classical resources for qubit control and error correction, directly impacting the performance of end-to-end quantum applications.

\emph{(ii) Heterogeneous Computing:} As quantum computing evolves towards integrating multiple quantum processors (QPU) with classical accelerators like FPGA and GPUs, quantum mini-apps can help estimate common workload and resource patterns. This is crucial for evaluating quantum software systems that can handle the increased complexity and heterogeneity of high-performance computing (HPC) environments and their integration with QPUs.

\emph{(ii) Scalability:} Quantum mini-apps can help design and evaluate quantum software with a holistic, end-to-end perspective focused on the scalability of every stack layer. This approach is essential as it ensures that the software can grow alongside the evolving quantum hardware landscape without requiring complete redesigns for each technological advancement.

\subsection{Mini-App Framework}
\begin{figure}
  \includegraphics[width=0.40\textwidth]{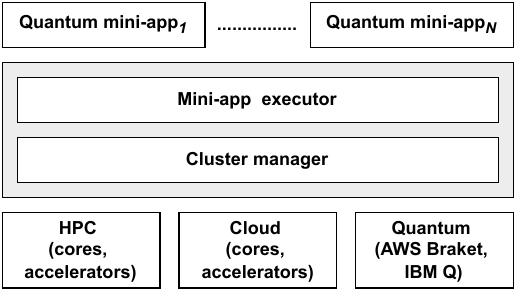}
  \caption{Quantum Mini-App Framework Architecture:  The framework facilitates the creation, execution, and management of quantum mini-apps. It comprises a mini-app executor and cluster manager abstracting different computing environments, including HPC, cloud, and quantum resources. \label{fig:qc_mini_app_arch}}

  \label{fig:mini-app-arch}
\end{figure}

Figure~\ref{fig:mini-app-arch} shows the architecture of the quantum mini-app framework. The framework serves as a basis for creating, implementing, running, and analyzing high-level motifs of a mini-app across distributed resources. It is designed to be flexible and scalable, enabling the creation and integration of new quantum mini-apps encompassing one or motifs. Mini-apps can be designed to be highly configurable and customizable, e.\,g., qubit count, circuit depth, and supported backends.

The mini-app framework comprises the \emph{mini-app executor} and the \emph{cluster manager}. The \emph{mini-app executor} orchestrates the execution of a mini-app across distributed computing resources. It manages allocating tasks to the available resources, ensuring efficient utilization of the computational power, and aggregates the results obtained from executing these tasks. 

The \emph{cluster manager} provides an interface to interact with clusters and is extensible to provision clusters for managing quantum/classical tasks across local, HPC, and cloud-based CPUs, GPUs, and QPUs. Currently, the mini-app executor supports the \emph{Dask} engine~\cite{dask}. Still, its modular architecture allows seamless integration with other engines, such as \emph{Ray}, to benchmark the specific requirements of different quantum computing applications. For example, \emph{Ray} supports the inherent mapping of tasks to GPUs, while Dask doesn't.

\subsection{Mini-App Examples}
We propose three mini-apps involving one or more motifs that isolate specific aspects of larger quantum-HPC applications. 

\subsubsection{Quantum Simulation}

The Quantum Simulation mini-app implements the \emph{circuit execution} motif utilizing the Qiskit library~\cite{qiskit_1} for generating random quantum circuits and executes these using different Aer simulator backends, e.\,g., with and without GPU support. We utilize a Dask distributed cluster environment managed via the mini-app framework to manage tasks across multiple nodes.

The distributed state vector motif implementation is based on the Pennylane lightning simulator~\cite{asadi2024hybrid}. The lightning simulator utilizes different HPC techniques, e.\,g., MPI, cuQuantum, and Kokkos provide a scalable, to provide a scalable distributed state vector simulation. It can simulate larger qubit counts and circuit depths. As the number of qubits to simulate increases, the growth of memory requirements and the need for scaling the simulation across multiple GPUs and nodes increases significantly.

\subsubsection{Variational Quantum Algorithms}
Variational algorithms represent, as described, an important class of NISQ algorithms that leverage the interplay between quantum and classical tasks~\cite{Cerezo_2021}. Typically, VQA comprises three key components: a parameterized quantum circuit, a classical optimizer, and a cost function. Variational circuits are trained iteratively within a classic optimization loop, which involves adjusting the parameters to approximate the ground state with a minimized cost function. The cost function typically depends upon the type of algorithm (i.\,e., VQE, QAOA, GAN). The arrangement of these components can vary, e.\,g., the quantum and classical components can be concurrently executed. Further, quantum components could be executed on multiple QPUs (including simulated QPUs).

\subsubsection{QML Pipeline} 

The QML mini-app provides a concrete example for the \textit{multistage pipeline} motif. We represent the QML pipelines mini-app into two stages. The first stage provides a standardized quantum representation of classical data. In contrast, the second stage subsequently assesses the performance of quantum and classical classifiers on the standardized datasets (e.\,g., MNIST~\cite{deng2012mnist} and ImageNet~\cite{deng2009imagenet}). In the first stage, the QML pipeline mini-app can find the approximate quantum representation of classical data mapped to a quantum state using techniques such as matrix product states (MPS)~\cite{Or_s_2014}. After fitting the MPS to classical data points, it can be translated to its quantum circuit representation. Since fitting the quantum circuits to the individual data points is independent, the optimizations can be run in parallel. After encoding the whole dataset, the QML pipeline mini-app can be utilized to assess the performance of classical and quantum classifiers in the second stage.\\
\\
As a proof-of-concept, we provide an implementation of the quantum simulation mini-app as part of the current framework. In the following, we investigate the performance of this mini-app.

\section{Experimental Results}
\label{sec:eval}

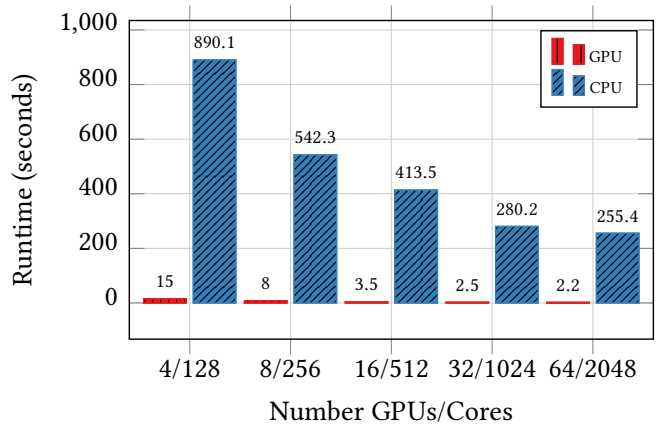
\begin{figure}[t]
  \centering
  \resizebox{0.5\textwidth}{!}{%
    \begin{tikzpicture}

\begin{axis}[legend style={nodes={scale=0.6, transform shape}}, 
        legend cell align=right,legend style={legend pos=north east,font=\small},
ybar,
bar width=12pt,
enlargelimits=0.15,
ylabel={Runtime (seconds)},
xlabel={Number GPUs/Cores},
symbolic x coords={4/128,8/256,16/512,32/1024,64/2048},
xtick=data,
nodes near coords,
every node near coord/.append style={font=\tiny},
ymax=900,
ytick={0,200,...,1000},
nodes near coords,
xticklabel style = {font=\small,yshift=0.5ex},
yticklabel style = {font=\small,yshift=0.5ex},
tick label style={font=\tiny},   
height=135pt,
width=0.8\linewidth,
cycle list name=set2,
grid = both,
grid style={line width=.1pt, draw=gray!40},
label style={font=\small} 
]
\addplot+[
    ybar, text=black,
    postaction={
        pattern=vertical lines
    }
] coordinates {(4/128, 15.0) (8/256, 8.0) (16/512, 3.5) (32/1024, 2.5) (64/2048, 2.2)};

\addplot+[
    ybar, text=black,
    postaction={
        pattern=north east lines
    }
] coordinates {(4/128, 890.1) (8/256, 542.3) (16/512, 413.5) (32/1024, 280.2) (64/2048, 255.4)};

\legend{GPU, CPU}
\end{axis}
\end{tikzpicture}
  }
  \caption{\textbf{Circuit execution motif characterization on Perlmutter: compute time for 1024 randomized circuit tasks for 25 qubits}. Executed 1024 circuit execution tasks on up to 16 Perlmutter GPU \& CPU nodes, where each GPU node has 4 A100 GPUs and CPU node has 128 cores. We scale from 1~node (4/128 GPUs/Cores) to 16 nodes (64/2048 GPUs/Cores). }
  \label{fig:qiskit_ce_scaling}
\end{figure}

In this section, we demonstrate the framework using an implementation of quantum simulation example mini-app focusing on the circuit execution motif with $1024$ randomly generated quantum circuits, \textit{qubits} size of $25$ and a circuit \textit{depth} of $1$.  We utilized Qiskit's \textit{random\_circuit module} to generate the circuits and observables to run state vector simulation using Qiskit's Aer simulator with cuQuantum-support.  We performed our experiments on the Perlmutter machine at NERSC\footnote{www.docs.nersc.gov/systems/perlmutter}. Perlmutter comprises $3072$ CPU-only and $1792$ GPU-accelerated nodes. Each CPU-only node has two CPUs with $64$ cores each, i.\,e., a total of $128$ cores. Each GPU node has a single CPU with $64$ cores and four $A100$ GPUs with $\SI{40}{\giga\byte}$ of GPU memory. We utilize up to 16 nodes, comparing CPU and GPU performance.

As the number of GPUs and cores increases, there is a clear decreasing trend for both CPU and GPU scenarios. Increasing the number of GPUs from 4 to 64 resulted in an $85\%$ reduction in runtime.  In contrast, CPU runtimes start significantly higher at $890\,sec$ seconds for $1$ node with $128$ cores decreasing to $255\,sec$ for $2048$ cores, a $71\%$ enhancement. The experimental data clearly illustrates the superior efficiency of GPUs. Across all experiments, GPU runs are approximately $95$ times faster. 

The quantum simulation mini-app facilitates benchmarking of important application circuits across diverse quantum-HPC infrastructures, enabling comparisons across various configurations. Although the current focus is on classically simulating quantum circuits, the mini-app framework can be easily extended to physical QPUs. Mini-apps like this are pivotal in advancing quantum applications, algorithms, and infrastructure. Insights from these experiments can contribute to the development and refinement of both software and hardware infrastructure, such as improving simulation capabilities and enhancing middleware functionalities.

\section{Related Work}
\label{sec:related}

Mini-apps can be used as proxies for real applications and workflows to enhance analysis and improve performance. Mini-apps isolate specific aspects of larger applications, allowing efficient exploration of hardware and software configurations. The concept is broadly used for testing, characterizing, and benchmarking application performance in various computing environments, e.\,g.,  data analytics~\cite{Sukumar2016Miniapps}, streaming~\cite{pilot-streaming}, and workflows~\cite{workflow_miniapps_jha2024}. 

Several mini-apps frameworks, such as  ECP proxy apps~\cite{ang2020ecp}, ECP CANDLE Benchmarks~\cite{wu2019performance}, ExaMiniMD~\cite{crozier2009improving}, and miniVite~\cite{ghosh2018minivite}, exist in the HPC domain, which are more application-specific. Extending upon such works, NERSC-10 Benchmark Suite~\cite{nerscNERSC10Benchmark} provides application-agnostic micro-benchmarks for evaluating the performance of different tasks within one or more application workflows. Other initiatives such as Neural mini-apps~\cite{vineyard2022neural}, and Neuromapp~\cite{ewart2017neuromapp} also emerged to understand the performance of emerging computing paradigms such as neuromorphic computing and improvement of neural simulators, respectively.

Recently, many researchers have focused on developing quantum benchmarks. System-level performance benchmarks such as QV~\cite{cross2019validating} and VB~\cite{blume2020volumetric} measure quantum circuit execution quality and scale (i.\,e., using qubit number and gate fidelity). At the same time, CLOPS~\cite{wack2021quality} characterizes runtime performance and execution time using quantum volume circuit sequence execution. Application-oriented quantum computing benchmarks, such as QPack~\cite{koen_qaoa_benchmarking_2022}, QED-C~\cite{Lubinski_etal_2023, Lubinski_etal_2021,lubinski2024quantum} and Q-Score~\cite{Martiel_2021} enable characterizing the performance of individual circuit execution (i.\,e. execution quality, runtime, gate count, etc.) for quantum applications of a specific type across diverse quantum hardware and simulators. While the above benchmarks are application-specific, the  QUARK framework~\cite{Finzgar_2022,kiwit2023application,kiwit2024benchmarking} orchestrates different application-oriented quantum benchmarks in a well-defined, standardized, reproducible, and verifiable way. It also incorporates workloads from the domains of optimization and quantum machine learning. As a future work, we aim to integrate and utilize a QUARK-like framework with quantum mini-apps to generalize benchmarking of high-level execution motifs composed within diverse mini-apps. 

\section{Conclusion}
\label{sec:conclusion}

Quantum-HPC execution motifs presented in this paper provide a conceptual understanding of the quantum-classical applications and workloads. These motifs provide the conceptual underpinning for \emph{quantum mini-app}. The framework and quantum simulation mini-app are open source and available on Github~\cite{QuantumMiniApp}. Further, the mini-app framework serves as the basis for implementing application kernels that represent these motifs, offering a structured approach for exploring the scalability and performance of quantum applications. Further, it provides the basis for developing advanced middleware systems.

In future work, we aim to develop and implement additional quantum mini-apps, e.\,g, for variational algorithms and QML pipe\-lines, to represent the six identified motifs comprehensively. These mini-apps will be critical tools for systematically exploring and refining the interactions between quantum hardware and middleware. Furthermore, we aim to develop a middleware system named \emph{pilot-quantum} based on the pilot abstraction~\cite{6404423}. This system will simplify resource and workload management for quantum-HPC systems by abstracting the allocation, management, and optimization of heterogeneous resources and tasks.

\section*{Acknowledgment}
This work received support from NERSC (Project ERCAP0029512), ORNL (Project CSC595), and the Bavarian State Ministry of Economic Affairs (BenchQC project, Grant DIK-0425/03). The authors generated parts of this text using OpenAI's language-generation models. Upon generation, the authors reviewed, edited, and revised the language.

\bibliographystyle{ACM-Reference-Format}
\bibliography{bibliography.bib}


\begin{thebibliography}{66}


\ifx \showCODEN    \undefined \def \showCODEN     #1{\unskip}     \fi
\ifx \showDOI      \undefined \def \showDOI       #1{#1}\fi
\ifx \showISBNx    \undefined \def \showISBNx     #1{\unskip}     \fi
\ifx \showISBNxiii \undefined \def \showISBNxiii  #1{\unskip}     \fi
\ifx \showISSN     \undefined \def \showISSN      #1{\unskip}     \fi
\ifx \showLCCN     \undefined \def \showLCCN      #1{\unskip}     \fi
\ifx \shownote     \undefined \def \shownote      #1{#1}          \fi
\ifx \showarticletitle \undefined \def \showarticletitle #1{#1}   \fi
\ifx \showURL      \undefined \def \showURL       {\relax}        \fi
\providecommand\bibfield[2]{#2}
\providecommand\bibinfo[2]{#2}
\providecommand\natexlab[1]{#1}
\providecommand\showeprint[2][]{arXiv:#2}

\bibitem[Alcazar et~al\mbox{.}(2021)]%
        {https://doi.org/10.48550/arxiv.2101.06250}
\bibfield{author}{\bibinfo{person}{Javier Alcazar},
  \bibinfo{person}{Mohammad~Ghazi Vakili}, \bibinfo{person}{Can~B. Kalayci},
  {and} \bibinfo{person}{Alejandro Perdomo-Ortiz}.}
  \bibinfo{year}{2021}\natexlab{}.
\newblock \bibinfo{title}{GEO: Enhancing Combinatorial Optimization with
  Classical and Quantum Generative Models}.
\newblock
\newblock
\urldef\tempurl%
\url{https://doi.org/10.48550/ARXIV.2101.06250}
\showDOI{\tempurl}


\bibitem[Alexeev et~al\mbox{.}(2023)]%
        {alexeev2023quantumcentric}
\bibfield{author}{\bibinfo{person}{Yuri Alexeev}, \bibinfo{person}{Maximilian
  Amsler}, \bibinfo{person}{Paul Baity}, \bibinfo{person}{Marco~Antonio
  Barroca}, \bibinfo{person}{Sanzio Bassini}, \bibinfo{person}{Torey Battelle},
  \bibinfo{person}{Daan Camps}, \bibinfo{person}{David Casanova},
  \bibinfo{person}{Young jai Choi}, \bibinfo{person}{Frederic~T. Chong},
  \bibinfo{person}{Charles Chung}, \bibinfo{person}{Chris Codella},
  \bibinfo{person}{Antonio~D. Corcoles}, \bibinfo{person}{James Cruise},
  \bibinfo{person}{Alberto~Di Meglio}, \bibinfo{person}{Jonathan Dubois},
  \bibinfo{person}{Ivan Duran}, \bibinfo{person}{Thomas Eckl},
  \bibinfo{person}{Sophia Economou}, \bibinfo{person}{Stephan Eidenbenz},
  \bibinfo{person}{Bruce Elmegreen}, \bibinfo{person}{Clyde Fare},
  \bibinfo{person}{Ismael Faro}, \bibinfo{person}{Cristina~Sanz Fernández},
  \bibinfo{person}{Rodrigo Neumann~Barros Ferreira}, \bibinfo{person}{Keisuke
  Fuji}, \bibinfo{person}{Bryce Fuller}, \bibinfo{person}{Laura Gagliardi},
  \bibinfo{person}{Giulia Galli}, \bibinfo{person}{Jennifer~R. Glick},
  \bibinfo{person}{Isacco Gobbi}, \bibinfo{person}{Pranav Gokhale},
  \bibinfo{person}{Salvador de~la Puente~Gonzalez}, \bibinfo{person}{Johannes
  Greiner}, \bibinfo{person}{Bill Gropp}, \bibinfo{person}{Michele Grossi},
  \bibinfo{person}{Emmanuel Gull}, \bibinfo{person}{Burns Healy},
  \bibinfo{person}{Benchen Huang}, \bibinfo{person}{Travis~S. Humble},
  \bibinfo{person}{Nobuyasu Ito}, \bibinfo{person}{Artur~F. Izmaylov},
  \bibinfo{person}{Ali Javadi-Abhari}, \bibinfo{person}{Douglas Jennewein},
  \bibinfo{person}{Shantenu Jha}, \bibinfo{person}{Liang Jiang},
  \bibinfo{person}{Barbara Jones}, \bibinfo{person}{Wibe~Albert de Jong},
  \bibinfo{person}{Petar Jurcevic}, \bibinfo{person}{William Kirby},
  \bibinfo{person}{Stefan Kister}, \bibinfo{person}{Masahiro Kitagawa},
  \bibinfo{person}{Joel Klassen}, \bibinfo{person}{Katherine Klymko},
  \bibinfo{person}{Kwangwon Koh}, \bibinfo{person}{Masaaki Kondo},
  \bibinfo{person}{Doga~Murat Kurkcuoglu}, \bibinfo{person}{Krzysztof
  Kurowski}, \bibinfo{person}{Teodoro Laino}, \bibinfo{person}{Ryan Landfield},
  \bibinfo{person}{Matt Leininger}, \bibinfo{person}{Vicente Leyton-Ortega},
  \bibinfo{person}{Ang Li}, \bibinfo{person}{Meifeng Lin},
  \bibinfo{person}{Junyu Liu}, \bibinfo{person}{Nicolas Lorente},
  \bibinfo{person}{Andre Luckow}, \bibinfo{person}{Simon Martiel},
  \bibinfo{person}{Francisco Martin-Fernandez}, \bibinfo{person}{Margaret
  Martonosi}, \bibinfo{person}{Claire Marvinney},
  \bibinfo{person}{Arcesio~Castaneda Medina}, \bibinfo{person}{Dirk Merten},
  \bibinfo{person}{Antonio Mezzacapo}, \bibinfo{person}{Kristel Michielsen},
  \bibinfo{person}{Abhishek Mitra}, \bibinfo{person}{Tushar Mittal},
  \bibinfo{person}{Kyungsun Moon}, \bibinfo{person}{Joel Moore},
  \bibinfo{person}{Mario Motta}, \bibinfo{person}{Young-Hye Na},
  \bibinfo{person}{Yunseong Nam}, \bibinfo{person}{Prineha Narang},
  \bibinfo{person}{Yu ya Ohnishi}, \bibinfo{person}{Daniele Ottaviani},
  \bibinfo{person}{Matthew Otten}, \bibinfo{person}{Scott Pakin},
  \bibinfo{person}{Vincent~R. Pascuzzi}, \bibinfo{person}{Ed Penault},
  \bibinfo{person}{Tomasz Piontek}, \bibinfo{person}{Jed Pitera},
  \bibinfo{person}{Patrick Rall}, \bibinfo{person}{Gokul~Subramanian Ravi},
  \bibinfo{person}{Niall Robertson}, \bibinfo{person}{Matteo Rossi},
  \bibinfo{person}{Piotr Rydlichowski}, \bibinfo{person}{Hoon Ryu},
  \bibinfo{person}{Georgy Samsonidze}, \bibinfo{person}{Mitsuhisa Sato},
  \bibinfo{person}{Nishant Saurabh}, \bibinfo{person}{Vidushi Sharma},
  \bibinfo{person}{Kunal Sharma}, \bibinfo{person}{Soyoung Shin},
  \bibinfo{person}{George Slessman}, \bibinfo{person}{Mathias Steiner},
  \bibinfo{person}{Iskandar Sitdikov}, \bibinfo{person}{In-Saeng Suh},
  \bibinfo{person}{Eric Switzer}, \bibinfo{person}{Wei Tang},
  \bibinfo{person}{Joel Thompson}, \bibinfo{person}{Synge Todo},
  \bibinfo{person}{Minh Tran}, \bibinfo{person}{Dimitar Trenev},
  \bibinfo{person}{Christian Trott}, \bibinfo{person}{Huan-Hsin Tseng},
  \bibinfo{person}{Esin Tureci}, \bibinfo{person}{David~García Valinas},
  \bibinfo{person}{Sofia Vallecorsa}, \bibinfo{person}{Christopher Wever},
  \bibinfo{person}{Konrad Wojciechowski}, \bibinfo{person}{Xiaodi Wu},
  \bibinfo{person}{Shinjae Yoo}, \bibinfo{person}{Nobuyuki Yoshioka},
  \bibinfo{person}{Victor~Wen zhe Yu}, \bibinfo{person}{Seiji Yunoki},
  \bibinfo{person}{Sergiy Zhuk}, {and} \bibinfo{person}{Dmitry Zubarev}.}
  \bibinfo{year}{2023}\natexlab{}.
\newblock \bibinfo{title}{Quantum-centric Supercomputing for Materials Science:
  A Perspective on Challenges and Future Directions}.
\newblock
\newblock
\showeprint[arxiv]{2312.09733}~[quant-ph]


\bibitem[Ang et~al\mbox{.}(2020)]%
        {ang2020ecp}
\bibfield{author}{\bibinfo{person}{Jim Ang}, \bibinfo{person}{Christine
  Sweeney}, \bibinfo{person}{Michael Wolf}, \bibinfo{person}{John~Austin
  Ellis}, \bibinfo{person}{Sayan Ghosh}, \bibinfo{person}{Ai Kagawa},
  \bibinfo{person}{Yunzhi Huang}, \bibinfo{person}{Sivasankaran Rajamanickam},
  \bibinfo{person}{Vinay Ramakrishnaiah}, \bibinfo{person}{Malachi Schram},
  {et~al\mbox{.}}} \bibinfo{year}{2020}\natexlab{}.
\newblock \bibinfo{booktitle}{\emph{ECP report: Update on proxy applications
  and vendor interactions}}.
\newblock \bibinfo{type}{{T}echnical {R}eport}. \bibinfo{institution}{Sandia
  National Lab.(SNL-NM), Albuquerque, NM (United States)}.
\newblock


\bibitem[Asadi et~al\mbox{.}(2024)]%
        {asadi2024hybrid}
\bibfield{author}{\bibinfo{person}{Ali Asadi}, \bibinfo{person}{Amintor Dusko},
  \bibinfo{person}{Chae-Yeun Park}, \bibinfo{person}{Vincent Michaud-Rioux},
  \bibinfo{person}{Isidor Schoch}, \bibinfo{person}{Shuli Shu},
  \bibinfo{person}{Trevor Vincent}, {and} \bibinfo{person}{Lee~James
  O'Riordan}.} \bibinfo{year}{2024}\natexlab{}.
\newblock \bibinfo{title}{Hybrid quantum programming with PennyLane Lightning
  on HPC platforms}.
\newblock
\newblock
\showeprint[arxiv]{2403.02512}~[quant-ph]


\bibitem[Bayerstadler et~al\mbox{.}(2021)]%
        {bayerstadler2021industry}
\bibfield{author}{\bibinfo{person}{Andreas Bayerstadler},
  \bibinfo{person}{Guillaume Becquin}, \bibinfo{person}{Julia Binder},
  \bibinfo{person}{Thierry Botter}, \bibinfo{person}{Hans Ehm},
  \bibinfo{person}{Thomas Ehmer}, \bibinfo{person}{Marvin Erdmann},
  \bibinfo{person}{Norbert Gaus}, \bibinfo{person}{Philipp Harbach},
  \bibinfo{person}{Maximilian Hess}, {et~al\mbox{.}}}
  \bibinfo{year}{2021}\natexlab{}.
\newblock \showarticletitle{Industry quantum computing applications}.
\newblock \bibinfo{journal}{\emph{EPJ Quantum Technology}} \bibinfo{volume}{8},
  \bibinfo{number}{1} (\bibinfo{year}{2021}), \bibinfo{pages}{25}.
\newblock


\bibitem[Bello et~al\mbox{.}(2023)]%
        {circuit-knitting-toolbox}
\bibfield{author}{\bibinfo{person}{Luciano Bello}, \bibinfo{person}{Agata~M.
  Bra\'{n}czyk}, \bibinfo{person}{Sergey Bravyi}, \bibinfo{person}{Almudena
  {Carrera Vazquez}}, \bibinfo{person}{Andrew Eddins},
  \bibinfo{person}{Daniel~J. Egger}, \bibinfo{person}{Bryce Fuller},
  \bibinfo{person}{Julien Gacon}, \bibinfo{person}{James~R. Garrison},
  \bibinfo{person}{Jennifer~R. Glick}, \bibinfo{person}{Tanvi~P. Gujarati},
  \bibinfo{person}{Ikko Hamamura}, \bibinfo{person}{Areeq~I. Hasan},
  \bibinfo{person}{Takashi Imamichi}, \bibinfo{person}{Caleb Johnson},
  \bibinfo{person}{Ieva Liepuoniute}, \bibinfo{person}{Owen Lockwood},
  \bibinfo{person}{Mario Motta}, \bibinfo{person}{C.~D. Pemmaraju},
  \bibinfo{person}{Pedro Rivero}, \bibinfo{person}{Max Rossmannek},
  \bibinfo{person}{Travis~L. Scholten}, \bibinfo{person}{Seetharami Seelam},
  \bibinfo{person}{Iskandar Sitdikov}, \bibinfo{person}{Dharmashankar
  Subramanian}, \bibinfo{person}{Wei Tang}, {and} \bibinfo{person}{Stefan
  Woerner}.} \bibinfo{year}{2023}\natexlab{}.
\newblock \bibinfo{title}{{Circuit Knitting Toolbox}}.
\newblock
  \bibinfo{howpublished}{\url{https://github.com/Qiskit-Extensions/circuit-knitting-toolbox}}.
\newblock
\urldef\tempurl%
\url{https://doi.org/10.5281/zenodo.7987997}
\showDOI{\tempurl}


\bibitem[Bergholm et~al\mbox{.}(2018)]%
        {bergholm2018pennylane}
\bibfield{author}{\bibinfo{person}{Ville Bergholm}, \bibinfo{person}{Josh
  Izaac}, \bibinfo{person}{Maria Schuld}, \bibinfo{person}{Christian Gogolin},
  \bibinfo{person}{M~Sohaib Alam}, \bibinfo{person}{Shahnawaz Ahmed},
  \bibinfo{person}{Juan~Miguel Arrazola}, \bibinfo{person}{Carsten Blank},
  \bibinfo{person}{Alain Delgado}, \bibinfo{person}{Soran Jahangiri},
  {et~al\mbox{.}}} \bibinfo{year}{2018}\natexlab{}.
\newblock \showarticletitle{Pennylane: Automatic differentiation of hybrid
  quantum-classical computations}.
\newblock \bibinfo{journal}{\emph{arXiv:1811.04968}} (\bibinfo{year}{2018}).
\newblock


\bibitem[Blume-Kohout and Young(2020)]%
        {blume2020volumetric}
\bibfield{author}{\bibinfo{person}{Robin Blume-Kohout} {and}
  \bibinfo{person}{Kevin~C Young}.} \bibinfo{year}{2020}\natexlab{}.
\newblock \showarticletitle{A volumetric framework for quantum computer
  benchmarks}.
\newblock \bibinfo{journal}{\emph{Quantum}}  \bibinfo{volume}{4}
  (\bibinfo{year}{2020}), \bibinfo{pages}{362}.
\newblock


\bibitem[Bravo-Prieto et~al\mbox{.}(2019)]%
        {vqls}
\bibfield{author}{\bibinfo{person}{Carlos Bravo-Prieto}, \bibinfo{person}{Ryan
  LaRose}, \bibinfo{person}{M. Cerezo}, \bibinfo{person}{Yigit Subasi},
  \bibinfo{person}{Lukasz Cincio}, {and} \bibinfo{person}{Patrick~J. Coles}.}
  \bibinfo{year}{2019}\natexlab{}.
\newblock \bibinfo{title}{Variational Quantum Linear Solver}.
\newblock
\newblock
\urldef\tempurl%
\url{https://doi.org/10.48550/ARXIV.1909.05820}
\showDOI{\tempurl}


\bibitem[Brewer et~al\mbox{.}(2024)]%
        {Brewer2024AICoupledHPC}
\bibfield{author}{\bibinfo{person}{Wesley Brewer}, \bibinfo{person}{Ana
  Gainaru}, \bibinfo{person}{Frédéric Suter}, \bibinfo{person}{Feiyi Wang},
  \bibinfo{person}{Murali Emani}, {and} \bibinfo{person}{Shantenu Jha}.}
  \bibinfo{year}{2024}\natexlab{}.
\newblock \showarticletitle{AI-coupled HPC Workflow Applications, Middleware
  and Performance}.
\newblock  (\bibinfo{year}{2024}).
\newblock
\newblock
\shownote{Under review}.


\bibitem[Cerezo et~al\mbox{.}(2021a)]%
        {vqa}
\bibfield{author}{\bibinfo{person}{M. Cerezo}, \bibinfo{person}{Andrew
  Arrasmith}, \bibinfo{person}{Ryan Babbush}, \bibinfo{person}{Simon~C.
  Benjamin}, \bibinfo{person}{Suguru Endo}, \bibinfo{person}{Keisuke Fujii},
  \bibinfo{person}{Jarrod~R. McClean}, \bibinfo{person}{Kosuke Mitarai},
  \bibinfo{person}{Xiao Yuan}, \bibinfo{person}{Lukasz Cincio}, {and}
  \bibinfo{person}{Patrick~J. Coles}.} \bibinfo{year}{2021}\natexlab{a}.
\newblock \showarticletitle{Variational quantum algorithms}.
\newblock \bibinfo{journal}{\emph{Nature Reviews Physics}} \bibinfo{volume}{3},
  \bibinfo{number}{9} (\bibinfo{year}{2021}), \bibinfo{pages}{625--644}.
\newblock
\showISBNx{2522-5820}
\urldef\tempurl%
\url{https://doi.org/10.1038/s42254-021-00348-9}
\showDOI{\tempurl}


\bibitem[Cerezo et~al\mbox{.}(2021b)]%
        {Cerezo_2021}
\bibfield{author}{\bibinfo{person}{M. Cerezo}, \bibinfo{person}{Andrew
  Arrasmith}, \bibinfo{person}{Ryan Babbush}, \bibinfo{person}{Simon~C.
  Benjamin}, \bibinfo{person}{Suguru Endo}, \bibinfo{person}{Keisuke Fujii},
  \bibinfo{person}{Jarrod~R. McClean}, \bibinfo{person}{Kosuke Mitarai},
  \bibinfo{person}{Xiao Yuan}, \bibinfo{person}{Lukasz Cincio}, {and}
  \bibinfo{person}{Patrick~J. Coles}.} \bibinfo{year}{2021}\natexlab{b}.
\newblock \showarticletitle{Variational quantum algorithms}.
\newblock \bibinfo{journal}{\emph{Nature Reviews Physics}} \bibinfo{volume}{3},
  \bibinfo{number}{9} (\bibinfo{date}{Aug.} \bibinfo{year}{2021}),
  \bibinfo{pages}{625–644}.
\newblock
\showISSN{2522-5820}
\urldef\tempurl%
\url{https://doi.org/10.1038/s42254-021-00348-9}
\showDOI{\tempurl}


\bibitem[Chantzialexiou et~al\mbox{.}(2018)]%
        {pilot-streaming}
\bibfield{author}{\bibinfo{person}{Georgios Chantzialexiou},
  \bibinfo{person}{Andre Luckow}, {and} \bibinfo{person}{Shantenu Jha}.}
  \bibinfo{year}{2018}\natexlab{}.
\newblock \showarticletitle{Pilot-Streaming: A Stream Processing Framework for
  High-Performance Computing}. In \bibinfo{booktitle}{\emph{2018 IEEE 14th
  International Conference on e-Science (e-Science)}}.
  \bibinfo{pages}{177--188}.
\newblock
\urldef\tempurl%
\url{https://doi.org/10.1109/eScience.2018.00033}
\showDOI{\tempurl}


\bibitem[Community(2024)]%
        {qiskit_1}
\bibfield{author}{\bibinfo{person}{Qiskit Community}.}
  \bibinfo{year}{2024}\natexlab{}.
\newblock \bibinfo{title}{Qiskit: An Open-source Framework for Quantum
  Computing}.
\newblock \bibinfo{howpublished}{\url{https://github.com/Qiskit/qiskit}}.
\newblock


\bibitem[C\'orcoles et~al\mbox{.}(2021)]%
        {PhysRevLett.127.100501}
\bibfield{author}{\bibinfo{person}{A.~D. C\'orcoles}, \bibinfo{person}{Maika
  Takita}, \bibinfo{person}{Ken Inoue}, \bibinfo{person}{Scott Lekuch},
  \bibinfo{person}{Zlatko~K. Minev}, \bibinfo{person}{Jerry~M. Chow}, {and}
  \bibinfo{person}{Jay~M. Gambetta}.} \bibinfo{year}{2021}\natexlab{}.
\newblock \showarticletitle{Exploiting Dynamic Quantum Circuits in a Quantum
  Algorithm with Superconducting Qubits}.
\newblock \bibinfo{journal}{\emph{Phys. Rev. Lett.}}  \bibinfo{volume}{127}
  (\bibinfo{date}{Aug} \bibinfo{year}{2021}), \bibinfo{pages}{100501}.
\newblock
Issue 10.
\urldef\tempurl%
\url{https://doi.org/10.1103/PhysRevLett.127.100501}
\showDOI{\tempurl}


\bibitem[Cranganore et~al\mbox{.}(2024)]%
        {cranganore2024paving}
\bibfield{author}{\bibinfo{person}{Sandeep~Suresh Cranganore},
  \bibinfo{person}{Vincenzo~De Maio}, \bibinfo{person}{Ivona Brandic}, {and}
  \bibinfo{person}{Ewa Deelman}.} \bibinfo{year}{2024}\natexlab{}.
\newblock \bibinfo{title}{Paving the Way to Hybrid Quantum-Classical Scientific
  Workflows}.
\newblock
\newblock
\showeprint[arxiv]{2404.10389}~[cs.ET]


\bibitem[Cross et~al\mbox{.}(2019)]%
        {cross2019validating}
\bibfield{author}{\bibinfo{person}{Andrew~W Cross}, \bibinfo{person}{Lev~S
  Bishop}, \bibinfo{person}{Sarah Sheldon}, \bibinfo{person}{Paul~D Nation},
  {and} \bibinfo{person}{Jay~M Gambetta}.} \bibinfo{year}{2019}\natexlab{}.
\newblock \showarticletitle{Validating quantum computers using randomized model
  circuits}.
\newblock \bibinfo{journal}{\emph{Physical Review A}} \bibinfo{volume}{100},
  \bibinfo{number}{3} (\bibinfo{year}{2019}), \bibinfo{pages}{032328}.
\newblock


\bibitem[Crozier et~al\mbox{.}(2009)]%
        {crozier2009improving}
\bibfield{author}{\bibinfo{person}{Paul~Stewart Crozier},
  \bibinfo{person}{Heidi~K Thornquist}, \bibinfo{person}{Robert~W Numrich},
  \bibinfo{person}{Alan~B Williams}, \bibinfo{person}{Harold~Carter Edwards},
  \bibinfo{person}{Eric~Richard Keiter}, \bibinfo{person}{Mahesh Rajan},
  \bibinfo{person}{James~M Willenbring}, \bibinfo{person}{Douglas~W Doerfler},
  {and} \bibinfo{person}{Michael~Allen Heroux}.}
  \bibinfo{year}{2009}\natexlab{}.
\newblock \bibinfo{booktitle}{\emph{Improving performance via
  mini-applications.}}
\newblock \bibinfo{type}{{T}echnical {R}eport}. \bibinfo{institution}{Sandia
  National Laboratories (SNL), Albuquerque, NM, and Livermore, CA~…}.
\newblock


\bibitem[Dalzell et~al\mbox{.}(2023)]%
        {dalzell2023quantum}
\bibfield{author}{\bibinfo{person}{Alexander~M. Dalzell}, \bibinfo{person}{Sam
  McArdle}, \bibinfo{person}{Mario Berta}, \bibinfo{person}{Przemyslaw
  Bienias}, \bibinfo{person}{Chi-Fang Chen}, \bibinfo{person}{András Gilyén},
  \bibinfo{person}{Connor~T. Hann}, \bibinfo{person}{Michael~J. Kastoryano},
  \bibinfo{person}{Emil~T. Khabiboulline}, \bibinfo{person}{Aleksander Kubica},
  \bibinfo{person}{Grant Salton}, \bibinfo{person}{Samson Wang}, {and}
  \bibinfo{person}{Fernando G. S.~L. Brandão}.}
  \bibinfo{year}{2023}\natexlab{}.
\newblock \bibinfo{title}{Quantum algorithms: A survey of applications and
  end-to-end complexities}.
\newblock
\newblock
\showeprint[arxiv]{2310.03011}~[quant-ph]


\bibitem[{Dask Development Team}(2024)]%
        {dask}
\bibfield{author}{\bibinfo{person}{{Dask Development Team}}.}
  \bibinfo{year}{2024}\natexlab{}.
\newblock \bibinfo{title}{Dask: Scale the Python tools you love}.
\newblock \bibinfo{howpublished}{\url{https://www.dask.org}}.
\newblock
\newblock
\shownote{Accessed: Apr. 21, 2024}.


\bibitem[Deng et~al\mbox{.}(2009)]%
        {deng2009imagenet}
\bibfield{author}{\bibinfo{person}{Jia Deng}, \bibinfo{person}{Wei Dong},
  \bibinfo{person}{Richard Socher}, \bibinfo{person}{Li-Jia Li},
  \bibinfo{person}{Kai Li}, {and} \bibinfo{person}{Li Fei-Fei}.}
  \bibinfo{year}{2009}\natexlab{}.
\newblock \showarticletitle{Imagenet: A large-scale hierarchical image
  database}. In \bibinfo{booktitle}{\emph{2009 IEEE conference on computer
  vision and pattern recognition}}. Ieee, \bibinfo{pages}{248--255}.
\newblock


\bibitem[Deng(2012)]%
        {deng2012mnist}
\bibfield{author}{\bibinfo{person}{Li Deng}.} \bibinfo{year}{2012}\natexlab{}.
\newblock \showarticletitle{The mnist database of handwritten digit images for
  machine learning research}.
\newblock \bibinfo{journal}{\emph{IEEE Signal Processing Magazine}}
  \bibinfo{volume}{29}, \bibinfo{number}{6} (\bibinfo{year}{2012}),
  \bibinfo{pages}{141--142}.
\newblock


\bibitem[Ella et~al\mbox{.}(2023)]%
        {ella2023quantumclassical}
\bibfield{author}{\bibinfo{person}{Lior Ella}, \bibinfo{person}{Lorenzo
  Leandro}, \bibinfo{person}{Oded Wertheim}, \bibinfo{person}{Yoav Romach},
  \bibinfo{person}{Ramon Szmuk}, \bibinfo{person}{Yoel Knol},
  \bibinfo{person}{Nissim Ofek}, \bibinfo{person}{Itamar Sivan}, {and}
  \bibinfo{person}{Yonatan Cohen}.} \bibinfo{year}{2023}\natexlab{}.
\newblock \bibinfo{title}{Quantum-classical processing and benchmarking at the
  pulse-level}.
\newblock
\newblock
\showeprint[arxiv]{2303.03816}~[quant-ph]


\bibitem[Elsharkawy et~al\mbox{.}(2023)]%
        {elsharkawy2023integration}
\bibfield{author}{\bibinfo{person}{Amr Elsharkawy},
  \bibinfo{person}{Xiao-Ting~Michelle To}, \bibinfo{person}{Philipp Seitz},
  \bibinfo{person}{Yanbin Chen}, \bibinfo{person}{Yannick Stade},
  \bibinfo{person}{Manuel Geiger}, \bibinfo{person}{Qunsheng Huang},
  \bibinfo{person}{Xiaorang Guo}, \bibinfo{person}{Muhammad~Arslan Ansari},
  \bibinfo{person}{Christian~B. Mendl}, \bibinfo{person}{Dieter Kranzlmüller},
  {and} \bibinfo{person}{Martin Schulz}.} \bibinfo{year}{2023}\natexlab{}.
\newblock \bibinfo{title}{Integration of Quantum Accelerators with High
  Performance Computing -- A Review of Quantum Programming Tools}.
\newblock
\newblock
\showeprint[arxiv]{2309.06167}~[cs.ET]


\bibitem[Ewart et~al\mbox{.}(2017)]%
        {ewart2017neuromapp}
\bibfield{author}{\bibinfo{person}{Timoth{\'e}e Ewart}, \bibinfo{person}{Judit
  Planas}, \bibinfo{person}{Francesco Cremonesi}, \bibinfo{person}{Kai Langen},
  \bibinfo{person}{Felix Sch{\"u}rmann}, {and} \bibinfo{person}{Fabien
  Delalondre}.} \bibinfo{year}{2017}\natexlab{}.
\newblock \showarticletitle{Neuromapp: a mini-application framework to improve
  neural simulators}. In \bibinfo{booktitle}{\emph{High Performance Computing:
  32nd International Conference, ISC High Performance 2017, Frankfurt, Germany,
  June 18--22, 2017, Proceedings 32}}. Springer, \bibinfo{pages}{181--198}.
\newblock


\bibitem[Fang et~al\mbox{.}(2023)]%
        {cuquantum}
\bibfield{author}{\bibinfo{person}{Leo Fang}, \bibinfo{person}{ahehn nv},
  \bibinfo{person}{hhbayraktar}, {and} \bibinfo{person}{sam stanwyck}.}
  \bibinfo{year}{2023}\natexlab{}.
\newblock \bibinfo{booktitle}{\emph{NVIDIA/cuQuantum: cuQuantum Python
  v22.11.0.1}}.
\newblock
\urldef\tempurl%
\url{https://doi.org/10.5281/zenodo.7523366}
\showDOI{\tempurl}


\bibitem[Farhi et~al\mbox{.}(2014)]%
        {farhi2014quantum}
\bibfield{author}{\bibinfo{person}{Edward Farhi}, \bibinfo{person}{Jeffrey
  Goldstone}, {and} \bibinfo{person}{Sam Gutmann}.}
  \bibinfo{year}{2014}\natexlab{}.
\newblock \showarticletitle{A quantum approximate optimization algorithm}.
\newblock \bibinfo{journal}{\emph{arXiv preprint arXiv:1411.4028}}
  (\bibinfo{year}{2014}).
\newblock


\bibitem[Finzgar et~al\mbox{.}(2022)]%
        {Finzgar_2022}
\bibfield{author}{\bibinfo{person}{Jernej~Rudi Finzgar},
  \bibinfo{person}{Philipp Ross}, \bibinfo{person}{Leonhard Holscher},
  \bibinfo{person}{Johannes Klepsch}, {and} \bibinfo{person}{Andre Luckow}.}
  \bibinfo{year}{2022}\natexlab{}.
\newblock \showarticletitle{QUARK: A Framework for Quantum Computing
  Application Benchmarking}. In \bibinfo{booktitle}{\emph{2022 IEEE
  International Conference on Quantum Computing and Engineering (QCE)}}.
  \bibinfo{publisher}{IEEE}.
\newblock
\urldef\tempurl%
\url{https://doi.org/10.1109/qce53715.2022.00042}
\showDOI{\tempurl}


\bibitem[Finžgar et~al\mbox{.}(2024)]%
        {finžgar2024quantuminformed}
\bibfield{author}{\bibinfo{person}{Jernej~Rudi Finžgar}, \bibinfo{person}{Aron
  Kerschbaumer}, \bibinfo{person}{Martin J.~A. Schuetz},
  \bibinfo{person}{Christian~B. Mendl}, {and} \bibinfo{person}{Helmut~G.
  Katzgraber}.} \bibinfo{year}{2024}\natexlab{}.
\newblock \bibinfo{title}{Quantum-Informed Recursive Optimization Algorithms}.
\newblock
\newblock
\showeprint[arxiv]{2308.13607}~[quant-ph]


\bibitem[Gebauer et~al\mbox{.}(2022)]%
        {generative_molecule_design_2022}
\bibfield{author}{\bibinfo{person}{Niklas W.~A. Gebauer},
  \bibinfo{person}{Michael Gastegger}, \bibinfo{person}{Stefaan S.~P.
  Hessmann}, \bibinfo{person}{Klaus-Robert M{\"u}ller}, {and}
  \bibinfo{person}{Kristof~T. Sch{\"u}tt}.} \bibinfo{year}{2022}\natexlab{}.
\newblock \showarticletitle{Inverse design of 3d molecular structures with
  conditional generative neural networks}.
\newblock \bibinfo{journal}{\emph{Nature Communications}} \bibinfo{volume}{13},
  \bibinfo{number}{1} (\bibinfo{year}{2022}), \bibinfo{pages}{973}.
\newblock
\showISBNx{2041-1723}
\urldef\tempurl%
\url{https://doi.org/10.1038/s41467-022-28526-y}
\showDOI{\tempurl}


\bibitem[Ghosh et~al\mbox{.}(2018)]%
        {ghosh2018minivite}
\bibfield{author}{\bibinfo{person}{Sayan Ghosh}, \bibinfo{person}{Mahantesh
  Halappanavar}, \bibinfo{person}{Antonino Tumeo}, \bibinfo{person}{Ananth
  Kalyanaraman}, {and} \bibinfo{person}{Assefaw~H Gebremedhin}.}
  \bibinfo{year}{2018}\natexlab{}.
\newblock \showarticletitle{miniVite: A graph analytics benchmarking tool for
  massively parallel systems}. In \bibinfo{booktitle}{\emph{2018 IEEE/ACM
  Performance Modeling, Benchmarking and Simulation of High Performance
  Computer Systems (PMBS)}}. IEEE, \bibinfo{pages}{51--56}.
\newblock


\bibitem[Imamura et~al\mbox{.}(2022)]%
        {imamura2022mpiqulacs}
\bibfield{author}{\bibinfo{person}{Satoshi Imamura}, \bibinfo{person}{Masafumi
  Yamazaki}, \bibinfo{person}{Takumi Honda}, \bibinfo{person}{Akihiko Kasagi},
  \bibinfo{person}{Akihiro Tabuchi}, \bibinfo{person}{Hiroshi Nakao},
  \bibinfo{person}{Naoto Fukumoto}, {and} \bibinfo{person}{Kohta Nakashima}.}
  \bibinfo{year}{2022}\natexlab{}.
\newblock \bibinfo{title}{mpiQulacs: A Distributed Quantum Computer Simulator
  for A64FX-based Cluster Systems}.
\newblock
\newblock
\showeprint[arxiv]{2203.16044}~[cs.DC]


\bibitem[Jha et~al\mbox{.}(2024)]%
        {jha2024aihpc}
\bibfield{author}{\bibinfo{person}{Shantenu Jha}, \bibinfo{person}{Wesley
  Brewer}, \bibinfo{person}{Ana Gainaru}, \bibinfo{person}{Frédéric Suter},
  \bibinfo{person}{Feiyi Wang}, {and} \bibinfo{person}{Murali Emani}.}
  \bibinfo{year}{2024}\natexlab{}.
\newblock \showarticletitle{Perspectives Paper: AI-HPC Workflow Applications,
  Middleware and Performance}.
\newblock  (\bibinfo{year}{2024}).
\newblock


\bibitem[Jobst et~al\mbox{.}(2023)]%
        {jobst2023efficient}
\bibfield{author}{\bibinfo{person}{Bernhard Jobst}, \bibinfo{person}{Kevin
  Shen}, \bibinfo{person}{Carlos~A. Riofrío}, \bibinfo{person}{Elvira
  Shishenina}, {and} \bibinfo{person}{Frank Pollmann}.}
  \bibinfo{year}{2023}\natexlab{}.
\newblock \bibinfo{title}{Efficient MPS representations and quantum circuits
  from the Fourier modes of classical image data}.
\newblock
\newblock
\showeprint[arxiv]{2311.07666}~[quant-ph]


\bibitem[Kilic et~al\mbox{.}(2024)]%
        {workflow_miniapps_jha2024}
\bibfield{author}{\bibinfo{person}{Ozgur~Ozan Kilic}, \bibinfo{person}{Tianle
  Wang}, \bibinfo{person}{Matteo Turilli}, \bibinfo{person}{Mikhail Titov},
  \bibinfo{person}{Andre Merzky}, \bibinfo{person}{Line Pouchard}, {and}
  \bibinfo{person}{Shantenu Jha}.} \bibinfo{year}{2024}\natexlab{}.
\newblock \bibinfo{title}{Workflow Mini-Apps: Portable, Scalable, Tunable \&
  Faithful Representations of Scientific Workflows}.
\newblock
\newblock
\showeprint[arxiv]{2403.18073}~[cs.DC]


\bibitem[Kiwit et~al\mbox{.}(2023)]%
        {kiwit2023application}
\bibfield{author}{\bibinfo{person}{Florian~J Kiwit}, \bibinfo{person}{Marwa
  Marso}, \bibinfo{person}{Philipp Ross}, \bibinfo{person}{Carlos~A
  Riofr{\'\i}o}, \bibinfo{person}{Johannes Klepsch}, {and}
  \bibinfo{person}{Andre Luckow}.} \bibinfo{year}{2023}\natexlab{}.
\newblock \showarticletitle{Application-Oriented Benchmarking of Quantum
  Generative Learning Using QUARK}. In \bibinfo{booktitle}{\emph{2023 IEEE
  International Conference on Quantum Computing and Engineering (QCE)}},
  Vol.~\bibinfo{volume}{1}. IEEE, \bibinfo{pages}{475--484}.
\newblock


\bibitem[Kiwit et~al\mbox{.}(2024)]%
        {kiwit2024benchmarking}
\bibfield{author}{\bibinfo{person}{Florian~J. Kiwit},
  \bibinfo{person}{Maximilian~A. Wolf}, \bibinfo{person}{Marwa Marso},
  \bibinfo{person}{Philipp Ross}, \bibinfo{person}{Jeanette~M. Lorenz},
  \bibinfo{person}{Carlos~A. Riofrío}, {and} \bibinfo{person}{Andre Luckow}.}
  \bibinfo{year}{2024}\natexlab{}.
\newblock \bibinfo{title}{Benchmarking Quantum Generative Learning: A Study on
  Scalability and Noise Resilience using QUARK}.
\newblock
\newblock
\showeprint[arxiv]{2403.18662}~[quant-ph]


\bibitem[Koen~Mesman(2022)]%
        {koen_qaoa_benchmarking_2022}
\bibfield{author}{\bibinfo{person}{Matthias~Möller Koen~Mesman, Zaid Al-Ars}.}
  \bibinfo{year}{2022}\natexlab{}.
\newblock \showarticletitle{QPack: Quantum Approximate Optimization Algorithms
  as universal benchmark for quantum computers}.
\newblock \bibinfo{journal}{\emph{arXiv}} (\bibinfo{year}{2022}).
\newblock
\urldef\tempurl%
\url{https://doi.org/10.48550/arXiv.2103.17193}
\showDOI{\tempurl}


\bibitem[Lubinski et~al\mbox{.}(2023a)]%
        {Lubinski_etal_2023}
\bibfield{author}{\bibinfo{person}{Thomas Lubinski}, \bibinfo{person}{Carleton
  Coffrin}, \bibinfo{person}{Catherine McGeoch}, \bibinfo{person}{Pratik
  Sathe}, \bibinfo{person}{Joshua Apanavicius}, {and} \bibinfo{person}{David
  E.~Bernal Neira}.} \bibinfo{year}{2023}\natexlab{a}.
\newblock \bibinfo{title}{Optimization Applications as Quantum Performance
  Benchmarks}.
\newblock
\newblock
\showeprint[arxiv]{2302.02278}~[quant-ph]


\bibitem[Lubinski et~al\mbox{.}(2024)]%
        {lubinski2024quantum}
\bibfield{author}{\bibinfo{person}{Thomas Lubinski}, \bibinfo{person}{Joshua~J.
  Goings}, \bibinfo{person}{Karl Mayer}, \bibinfo{person}{Sonika Johri},
  \bibinfo{person}{Nithin Reddy}, \bibinfo{person}{Aman Mehta},
  \bibinfo{person}{Niranjan Bhatia}, \bibinfo{person}{Sonny Rappaport},
  \bibinfo{person}{Daniel Mills}, \bibinfo{person}{Charles~H. Baldwin},
  \bibinfo{person}{Luning Zhao}, \bibinfo{person}{Aaron Barbosa},
  \bibinfo{person}{Smarak Maity}, {and} \bibinfo{person}{Pranav~S. Mundada}.}
  \bibinfo{year}{2024}\natexlab{}.
\newblock \bibinfo{title}{Quantum Algorithm Exploration using
  Application-Oriented Performance Benchmarks}.
\newblock
\newblock
\showeprint[arxiv]{2402.08985}~[quant-ph]


\bibitem[Lubinski et~al\mbox{.}(2023b)]%
        {Lubinski_etal_2021}
\bibfield{author}{\bibinfo{person}{Thomas Lubinski}, \bibinfo{person}{Sonika
  Johri}, \bibinfo{person}{Paul Varosy}, \bibinfo{person}{Jeremiah Coleman},
  \bibinfo{person}{Luning Zhao}, \bibinfo{person}{Jason Necaise},
  \bibinfo{person}{Charles~H. Baldwin}, \bibinfo{person}{Karl Mayer}, {and}
  \bibinfo{person}{Timothy Proctor}.} \bibinfo{year}{2023}\natexlab{b}.
\newblock \bibinfo{title}{Application-Oriented Performance Benchmarks for
  Quantum Computing}.
\newblock
\newblock
\showeprint[arxiv]{2110.03137}~[quant-ph]


\bibitem[Luckow and Jha(2019)]%
        {9006530}
\bibfield{author}{\bibinfo{person}{Andre Luckow} {and}
  \bibinfo{person}{Shantenu Jha}.} \bibinfo{year}{2019}\natexlab{}.
\newblock \showarticletitle{Performance Characterization and Modeling of
  Serverless and HPC Streaming Applications}. In \bibinfo{booktitle}{\emph{2019
  IEEE International Conference on Big Data (Big Data)}}.
  \bibinfo{pages}{5688--5696}.
\newblock
\urldef\tempurl%
\url{https://doi.org/10.1109/BigData47090.2019.9006530}
\showDOI{\tempurl}


\bibitem[Luckow et~al\mbox{.}(2012)]%
        {6404423}
\bibfield{author}{\bibinfo{person}{Andre Luckow}, \bibinfo{person}{Mark
  Santcroos}, \bibinfo{person}{Andre Merzky}, \bibinfo{person}{Ole Weidner},
  \bibinfo{person}{Pradeep Mantha}, {and} \bibinfo{person}{Shantenu Jha}.}
  \bibinfo{year}{2012}\natexlab{}.
\newblock \showarticletitle{P*: A model of pilot-abstractions}. In
  \bibinfo{booktitle}{\emph{2012 IEEE 8th International Conference on
  E-Science}}. \bibinfo{pages}{1--10}.
\newblock
\urldef\tempurl%
\url{https://doi.org/10.1109/eScience.2012.6404423}
\showDOI{\tempurl}


\bibitem[Mantha(2024)]%
        {QuantumMiniApp}
\bibfield{author}{\bibinfo{person}{Pradeep Mantha}.}
  \bibinfo{year}{2024}\natexlab{}.
\newblock \bibinfo{title}{Quantum Mini-App framework implementation}.
\newblock
  \bibinfo{howpublished}{\url{https://github.com/radical-cybertools/quantum-mini-apps}}.
\newblock
\newblock
\shownote{Accessed: 2024-03-31}.


\bibitem[Martiel et~al\mbox{.}(2021)]%
        {Martiel_2021}
\bibfield{author}{\bibinfo{person}{Simon Martiel}, \bibinfo{person}{Thomas
  Ayral}, {and} \bibinfo{person}{Cyril Allouche}.}
  \bibinfo{year}{2021}\natexlab{}.
\newblock \showarticletitle{Benchmarking Quantum Coprocessors in an
  Application-Centric, Hard\-ware-Agnostic, and Scalable Way}.
\newblock \bibinfo{journal}{\emph{IEEE Transactions on Quantum Engineering}}
  \bibinfo{volume}{2} (\bibinfo{year}{2021}), \bibinfo{pages}{1–11}.
\newblock
\showISSN{2689-1808}
\urldef\tempurl%
\url{https://doi.org/10.1109/tqe.2021.3090207}
\showDOI{\tempurl}


\bibitem[Nakaji et~al\mbox{.}(2024)]%
        {nakaji2024generative}
\bibfield{author}{\bibinfo{person}{Kouhei Nakaji},
  \bibinfo{person}{Lasse~Bjørn Kristensen}, \bibinfo{person}{Jorge~A.
  Campos-Gonzalez-Angulo}, \bibinfo{person}{Mohammad~Ghazi Vakili},
  \bibinfo{person}{Haozhe Huang}, \bibinfo{person}{Mohsen Bagherimehrab},
  \bibinfo{person}{Christoph Gorgulla}, \bibinfo{person}{FuTe Wong},
  \bibinfo{person}{Alex McCaskey}, \bibinfo{person}{Jin-Sung Kim},
  \bibinfo{person}{Thien Nguyen}, \bibinfo{person}{Pooja Rao}, {and}
  \bibinfo{person}{Alan Aspuru-Guzik}.} \bibinfo{year}{2024}\natexlab{}.
\newblock \bibinfo{title}{The generative quantum eigensolver (GQE) and its
  application for ground state search}.
\newblock
\newblock
\showeprint[arxiv]{2401.09253}~[quant-ph]


\bibitem[{National Academies of Sciences, Engineering, and Medicine}(2023)]%
        {NAP26850}
\bibfield{author}{\bibinfo{person}{{National Academies of Sciences,
  Engineering, and Medicine}}.} \bibinfo{year}{2023}\natexlab{}.
\newblock \bibinfo{booktitle}{\emph{Advancing Chemistry and Quantum Information
  Science: An Assessment of Research Opportunities at the Interface of
  Chemistry and Quantum Information Science in the United States}}.
\newblock \bibinfo{publisher}{The National Academies Press},
  \bibinfo{address}{Washington, DC}.
\newblock
\showISBNx{978-0-309-69809-2}
\urldef\tempurl%
\url{https://doi.org/10.17226/26850}
\showDOI{\tempurl}


\bibitem[{National Energy Research Scientific Computing Center (NERSC)}(2024)]%
        {nerscNERSC10Benchmark}
\bibfield{author}{\bibinfo{person}{{National Energy Research Scientific
  Computing Center (NERSC)}}.} \bibinfo{year}{2024}\natexlab{}.
\newblock \bibinfo{title}{{NERSC-10 Benchmark Suite --- nersc.gov}}.
\newblock
  \bibinfo{howpublished}{\url{https://www.nersc.gov/systems/nersc-10/benchmarks/}}.
\newblock
\newblock
\shownote{[Accessed 24-03-2024]}.


\bibitem[Orús(2014)]%
        {Or_s_2014}
\bibfield{author}{\bibinfo{person}{Román Orús}.}
  \bibinfo{year}{2014}\natexlab{}.
\newblock \showarticletitle{A practical introduction to tensor networks: Matrix
  product states and projected entangled pair states}.
\newblock \bibinfo{journal}{\emph{Annals of Physics}}  \bibinfo{volume}{349}
  (\bibinfo{date}{Oct.} \bibinfo{year}{2014}), \bibinfo{pages}{117–158}.
\newblock
\showISSN{0003-4916}
\urldef\tempurl%
\url{https://doi.org/10.1016/j.aop.2014.06.013}
\showDOI{\tempurl}


\bibitem[Peng et~al\mbox{.}(2020)]%
        {Peng_2020}
\bibfield{author}{\bibinfo{person}{Tianyi Peng}, \bibinfo{person}{Aram~W.
  Harrow}, \bibinfo{person}{Maris Ozols}, {and} \bibinfo{person}{Xiaodi Wu}.}
  \bibinfo{year}{2020}\natexlab{}.
\newblock \showarticletitle{Simulating Large Quantum Circuits on a Small
  Quantum Computer}.
\newblock \bibinfo{journal}{\emph{Physical Review Letters}}
  \bibinfo{volume}{125}, \bibinfo{number}{15} (\bibinfo{date}{Oct.}
  \bibinfo{year}{2020}).
\newblock
\showISSN{1079-7114}
\urldef\tempurl%
\url{https://doi.org/10.1103/physrevlett.125.150504}
\showDOI{\tempurl}


\bibitem[Peruzzo et~al\mbox{.}(2014)]%
        {vqe}
\bibfield{author}{\bibinfo{person}{Alberto Peruzzo}, \bibinfo{person}{Jarrod
  McClean}, \bibinfo{person}{Peter Shadbolt}, \bibinfo{person}{Man-Hong Yung},
  \bibinfo{person}{Xiao-Qi Zhou}, \bibinfo{person}{Peter~J. Love},
  \bibinfo{person}{Al{\'a}n Aspuru-Guzik}, {and} \bibinfo{person}{Jeremy~L.
  O'Brien}.} \bibinfo{year}{2014}\natexlab{}.
\newblock \showarticletitle{A variational eigenvalue solver on a photonic
  quantum processor}.
\newblock \bibinfo{journal}{\emph{Nature Communications}} \bibinfo{volume}{5},
  \bibinfo{number}{1} (\bibinfo{year}{2014}), \bibinfo{pages}{4213}.
\newblock
\showISBNx{2041-1723}
\urldef\tempurl%
\url{https://doi.org/10.1038/ncomms5213}
\showDOI{\tempurl}


\bibitem[Preskill(2018)]%
        {Preskill2018quantumcomputingin}
\bibfield{author}{\bibinfo{person}{John Preskill}.}
  \bibinfo{year}{2018}\natexlab{}.
\newblock \showarticletitle{Quantum {C}omputing in the {NISQ} era and beyond}.
\newblock \bibinfo{journal}{\emph{{Quantum}}}  \bibinfo{volume}{2}
  (\bibinfo{date}{Aug.} \bibinfo{year}{2018}), \bibinfo{pages}{79}.
\newblock
\showISSN{2521-327X}
\urldef\tempurl%
\url{https://doi.org/10.22331/q-2018-08-06-79}
\showDOI{\tempurl}


\bibitem[{Qiskit Development Team}(2023)]%
        {QiskitAerParallel}
\bibfield{author}{\bibinfo{person}{{Qiskit Development Team}}.}
  \bibinfo{year}{2023}\natexlab{}.
\newblock \bibinfo{title}{Running with Threadpool and DASK - Qiskit Aer
  0.13.1}.
\newblock
  \bibinfo{howpublished}{\url{https://qiskit.org/ecosystem/aer/howtos/parallel.html}}.
\newblock
\newblock
\shownote{Accessed: 2024-02-18}.


\bibitem[Quantum(2024)]%
        {IBMQuantumConfigureErrorMitigation}
\bibfield{author}{\bibinfo{person}{IBM Quantum}.}
  \bibinfo{year}{2024}\natexlab{}.
\newblock \bibinfo{title}{Configure error mitigation for Qiskit Runtime}.
\newblock
  \bibinfo{howpublished}{\url{https://docs.quantum.ibm.com/run/configure-error-mitigation}}.
\newblock
\newblock
\shownote{Accessed: 2024-02-18}.


\bibitem[{Quantum Technology and Application Consortium (QUTAC)}(2024)]%
        {QutacQuantumQuGen}
\bibfield{author}{\bibinfo{person}{{Quantum Technology and Application
  Consortium (QUTAC)}}.} \bibinfo{year}{2024}\natexlab{}.
\newblock \bibinfo{title}{{qugen: Quantum Circuit Generator}}.
\newblock \bibinfo{howpublished}{\url{https://github.com/QutacQuantum/qugen}}.
\newblock
\newblock
\shownote{Accessed: 2024-04-18}.


\bibitem[Ravi et~al\mbox{.}(2022)]%
        {ravi2022boosting}
\bibfield{author}{\bibinfo{person}{Gokul~Subramanian Ravi},
  \bibinfo{person}{Jonathan~M. Baker}, \bibinfo{person}{Kaitlin~N. Smith},
  \bibinfo{person}{Nathan Earnest}, \bibinfo{person}{Ali Javadi-Abhari}, {and}
  \bibinfo{person}{Frederic Chong}.} \bibinfo{year}{2022}\natexlab{}.
\newblock \bibinfo{title}{Boosting Quantum Fidelity with an Ordered Diverse
  Ensemble of Clifford Canary Circuits}.
\newblock
\newblock
\showeprint[arxiv]{2209.13732}~[quant-ph]


\bibitem[Riofr{\'\i}o et~al\mbox{.}(2023)]%
        {riofrio2023performance}
\bibfield{author}{\bibinfo{person}{Carlos~A Riofr{\'\i}o},
  \bibinfo{person}{Oliver Mitevski}, \bibinfo{person}{Caitlin Jones},
  \bibinfo{person}{Florian Krellner}, \bibinfo{person}{Aleksandar
  Vu{\v{c}}kovi{\'c}}, \bibinfo{person}{Joseph Doetsch},
  \bibinfo{person}{Johannes Klepsch}, \bibinfo{person}{Thomas Ehmer}, {and}
  \bibinfo{person}{Andre Luckow}.} \bibinfo{year}{2023}\natexlab{}.
\newblock \showarticletitle{A performance characterization of quantum
  generative models}.
\newblock \bibinfo{journal}{\emph{arXiv e-prints}} (\bibinfo{year}{2023}),
  \bibinfo{pages}{arXiv--2301}.
\newblock


\bibitem[Saurabh et~al\mbox{.}(2023)]%
        {saurabh2023conceptual}
\bibfield{author}{\bibinfo{person}{Nishant Saurabh}, \bibinfo{person}{Shantenu
  Jha}, {and} \bibinfo{person}{Andre Luckow}.} \bibinfo{year}{2023}\natexlab{}.
\newblock \showarticletitle{A Conceptual Architecture for a Quantum-HPC
  Middleware}. In \bibinfo{booktitle}{\emph{Proceedings of the IEEE Conference
  on Quantum Software}}. IEEE.
\newblock


\bibitem[Self et~al\mbox{.}(2021)]%
        {vqa_is}
\bibfield{author}{\bibinfo{person}{Chris~N. Self}, \bibinfo{person}{Kiran~E.
  Khosla}, \bibinfo{person}{Alistair W.~R. Smith},
  \bibinfo{person}{Fr{\'e}d{\'e}ric Sauvage}, \bibinfo{person}{Peter~D.
  Haynes}, \bibinfo{person}{Johannes Knolle}, \bibinfo{person}{Florian
  Mintert}, {and} \bibinfo{person}{M.~S. Kim}.}
  \bibinfo{year}{2021}\natexlab{}.
\newblock \showarticletitle{Variational quantum algorithm with information
  sharing}.
\newblock \bibinfo{journal}{\emph{npj Quantum Information}}
  \bibinfo{volume}{7}, \bibinfo{number}{1} (\bibinfo{year}{2021}),
  \bibinfo{pages}{116}.
\newblock
\showISBNx{2056-6387}
\urldef\tempurl%
\url{https://doi.org/10.1038/s41534-021-00452-9}
\showDOI{\tempurl}


\bibitem[Smith et~al\mbox{.}(2023)]%
        {10.1145/3579371.3589352}
\bibfield{author}{\bibinfo{person}{Kaitlin~N. Smith},
  \bibinfo{person}{Michael~A. Perlin}, \bibinfo{person}{Pranav Gokhale},
  \bibinfo{person}{Paige Frederick}, \bibinfo{person}{David Owusu-Antwi},
  \bibinfo{person}{Richard Rines}, \bibinfo{person}{Victory Omole}, {and}
  \bibinfo{person}{Frederic Chong}.} \bibinfo{year}{2023}\natexlab{}.
\newblock \showarticletitle{Clifford-based Circuit Cutting for Quantum
  Simulation}. In \bibinfo{booktitle}{\emph{Proceedings of the 50th Annual
  International Symposium on Computer Architecture}} (Orlando, FL, USA)
  \emph{(\bibinfo{series}{ISCA '23})}. \bibinfo{publisher}{Association for
  Computing Machinery}, \bibinfo{address}{New York, NY, USA}, Article
  \bibinfo{articleno}{84}, \bibinfo{numpages}{13}~pages.
\newblock
\showISBNx{9798400700958}
\urldef\tempurl%
\url{https://doi.org/10.1145/3579371.3589352}
\showDOI{\tempurl}


\bibitem[Sukumar et~al\mbox{.}(2016)]%
        {Sukumar2016Miniapps}
\bibfield{author}{\bibinfo{person}{S.~R. Sukumar}, \bibinfo{person}{M.~A.
  Matheson}, \bibinfo{person}{R. Kannan}, {and} \bibinfo{person}{S.-H. Lim}.}
  \bibinfo{year}{2016}\natexlab{}.
\newblock \showarticletitle{Mini-apps for High Performance Data Analysis}. In
  \bibinfo{booktitle}{\emph{2016 IEEE International Conference on Big Data (Big
  Data)}}. \bibinfo{publisher}{IEEE}, \bibinfo{pages}{1483--1492}.
\newblock


\bibitem[Tomesh et~al\mbox{.}(2023)]%
        {10313799}
\bibfield{author}{\bibinfo{person}{Teague Tomesh}, \bibinfo{person}{Zain~H.
  Saleem}, \bibinfo{person}{Michael~A. Perlin}, \bibinfo{person}{Pranav
  Gokhale}, \bibinfo{person}{Martin Suchara}, {and} \bibinfo{person}{Margaret
  Martonosi}.} \bibinfo{year}{2023}\natexlab{}.
\newblock \showarticletitle{Divide and Conquer for Combinatorial Optimization
  and Distributed Quantum Computation}. In \bibinfo{booktitle}{\emph{2023 IEEE
  International Conference on Quantum Computing and Engineering (QCE)}},
  Vol.~\bibinfo{volume}{01}. \bibinfo{pages}{1--12}.
\newblock
\urldef\tempurl%
\url{https://doi.org/10.1109/QCE57702.2023.00009}
\showDOI{\tempurl}


\bibitem[Vineyard et~al\mbox{.}(2022)]%
        {vineyard2022neural}
\bibfield{author}{\bibinfo{person}{Craig Vineyard}, \bibinfo{person}{Suma
  Cardwell}, \bibinfo{person}{Frances Chance}, \bibinfo{person}{Srideep
  Musuvathy}, \bibinfo{person}{Fred Rothganger}, \bibinfo{person}{William
  Severa}, \bibinfo{person}{John Smith}, \bibinfo{person}{Corinne Teeter},
  \bibinfo{person}{Felix Wang}, {and} \bibinfo{person}{James Aimone}.}
  \bibinfo{year}{2022}\natexlab{}.
\newblock \showarticletitle{Neural Mini-Apps as a Tool for Neuromorphic
  Computing Insight}. In \bibinfo{booktitle}{\emph{Proceedings of the 2022
  Annual Neuro-Inspired Computational Elements Conference}}.
  \bibinfo{pages}{40--49}.
\newblock


\bibitem[Wack et~al\mbox{.}(2021)]%
        {wack2021quality}
\bibfield{author}{\bibinfo{person}{Andrew Wack}, \bibinfo{person}{Hanhee Paik},
  \bibinfo{person}{Ali Javadi-Abhari}, \bibinfo{person}{Petar Jurcevic},
  \bibinfo{person}{Ismael Faro}, \bibinfo{person}{Jay~M. Gambetta}, {and}
  \bibinfo{person}{Blake~R. Johnson}.} \bibinfo{year}{2021}\natexlab{}.
\newblock \bibinfo{title}{Quality, Speed, and Scale: three key attributes to
  measure the performance of near-term quantum computers}.
\newblock
\newblock
\showeprint[arxiv]{2110.14108}~[quant-ph]


\bibitem[Wu et~al\mbox{.}(2019)]%
        {wu2019performance}
\bibfield{author}{\bibinfo{person}{Xingfu Wu}, \bibinfo{person}{Valerie
  Taylor}, \bibinfo{person}{Justin~M Wozniak}, \bibinfo{person}{Rick Stevens},
  \bibinfo{person}{Thomas Brettin}, {and} \bibinfo{person}{Fangfang Xia}.}
  \bibinfo{year}{2019}\natexlab{}.
\newblock \showarticletitle{Performance, energy, and scalability analysis and
  improvement of parallel cancer deep learning candle benchmarks}. In
  \bibinfo{booktitle}{\emph{Proceedings of the 48th International Conference on
  Parallel Processing}}. \bibinfo{pages}{1--11}.
\newblock


\bibitem[Xanadu(2023)]%
        {PennyLaneQCut}
\bibfield{author}{\bibinfo{person}{Xanadu}.} \bibinfo{year}{2023}\natexlab{}.
\newblock \bibinfo{title}{qml.qcut — PennyLane Documentation}.
\newblock
  \bibinfo{howpublished}{\url{https://docs.pennylane.ai/en/stable/code/qml_qcut.html}}.
\newblock
\newblock
\shownote{Accessed: 2024-02-18}.


\end{thebibliography}
\end{document}